\documentclass[12pt]{article}
\usepackage{times}
\usepackage{w-thm}
\usepackage[english]{babel}
\usepackage[latin1]{inputenc}

\usepackage{amsmath}

\usepackage[]{graphicx}
\usepackage{amsmath}
\usepackage{amssymb}
\usepackage{geometry}
\usepackage{booktabs}
\usepackage{float}
\usepackage{placeins}
\usepackage{url}
\usepackage{enumitem}

\newcommand{\blind}{0}

\begin{document}

\def\spacingset#1{\renewcommand{\baselinestretch}%
{#1}\small\normalsize} \spacingset{1}


\if0\blind
{
  \title{\bf A multiple comparison procedure for dose-finding trials with
subpopulations}
  \author{Marius Thomas\\
    Novartis Pharma AG, Basel, Switzerland\\
    \vspace{0.1cm}\\
    Bj\"orn Bornkamp\\ 
    Novartis Pharma AG, Basel, Switzerland\\
   \vspace{0.1cm}\\
    Martin Posch\\
    Medical University of Vienna, Vienna, Austria\\
   \vspace{0.1cm}\\
    Franz K\"onig\\
    Medical University of Vienna, Vienna, Austria}
  \date{}
  \maketitle
} \fi

\if1\blind
{
  \bigskip
  \bigskip
  \bigskip
  \begin{center}
    {\LARGE\bf A multiple comparison procedure for dose-finding trials with\\
subpopulations}
\end{center}
  \medskip
} \fi

\bigskip
\begin{abstract}
Identifying subgroups of patients with an enhanced response to a new treatment 
has become an area of increased interest in the last few years.
When there is knowledge about possible subpopulations with an enhanced
treatment effect before
the start of a trial it might be beneficial to set up a 
testing strategy, which tests for a significant treatment effect
not only in the full population, but also in these prespecified 
subpopulations.
In this paper we present a parametric multiple testing approach
for tests in multiple populations for dose-finding trials.
Our approach is based on the MCP-Mod methodology, 
which uses multiple comparison procedures to test for a 
dose-response signal, while considering 
multiple possible candidate dose-response shapes.
Our proposed methods allow for heteroscedasticity between populations
and control the FWER over tests in multiple
populations and for multiple candidate models. We show in simulations,
that the proposed multi-population testing approaches can increase
the power to detect a significant dose-response signal over
the standard single-population MCP-Mod, when the considered subpopulation
has an enhanced treatment effect.
\end{abstract}

\noindent%
\vfill

\noindent

\newpage
\spacingset{1.45} 

\section{Introduction}
\label{sec:intro}
Identifying subgroups of patients with an enhanced response to a new treatment
has become an area of increased interest in the last few years.
Hence subgroup analyses are commonly performed all throughout the
clinical development process.
In many cases these analyses are performed in an
exploratory fashion and a large number of possible subgroups are considered
with the aim to find a patient population, in which the treatment is
particularly effective. For these scenarios many novel subgroup
identification and treatment effect estimation methods have been 
proposed see, for example, 
\cite{fost:tayl:rube:2011, lipk:dmit:denn:2011, 
seib:zeil:hoth:2016, vara:wang:2016, 
rose:2016, shen:hu:li:2016} or \cite{liu:siva:laud:2017}.
An extensive overview over exploratory subgroup analysis methods
is given in \cite{lipk:dmit:agos:2017}.

When such an exploratory subgroup analysis results in a promising
subgroup finding, this subgroup could then be used
in a future trial. Alternatively such a subgroup could be identified
before the start of a trial based on clinical expert knowledge.
In these situation it might be beneficial
to plan the trial with the subgroup in mind and test
for a treatment effect in both the full population
and the subgroup. 
Depending on the result of the trial there are then three possible positive outcomes:
(i) the drug is shown to be efficacious for all patients, 
(ii) the drug is shown to be efficacious only in a subgroup or
(iii)  the drug is shown to be efficacious for all patients but with
an increased efficacy for patients in a subgroup. 
In the last scenario there might be uncertainty about the
effect in the full population, since it could be driven
by the large effect in the subgroup and the rest of the population
could have negligible treatment effects. One can consider
an additional test in the complement to avoid this issue.

For the design of trials, in which subpopulations
are of interest, additional
considerations have to be made regarding the power of the tests and
possible trial designs like enrichment or adaptive designs. Some
of these considerations regarding the power when testing in the whole population
and a subgroup are discussed in \cite{koch:1997} and \cite{alos:huqu:2009}.
\cite{ondr:dmit:frie:2016} discuss several different enrichment and
adaptive designs.

When tests are performed in multiple populations multiplicity
has to be taken in to account to avoid an inflated type I error.
It is possible to use non-parametric approaches to control multiplicity,
as for example, Bonferroni or Holm-corrections, or more complex
stepwise procedures as discussed in \cite{bret:maur:bran:2009}.
However non-parametric methods would ignore the correlations between
test statistics that arise, when tests are performed in the overall population
and a subgroup, where the latter is a subset of the former. 
Using a parametric approach can be more efficient and show 
an increase in power over the aforementioned methods.
Such parametric approaches are for example discussed in 
\cite{alos:huqu:2009} and \cite{bret:posc:glim:2011}.

In addition to testing in multiple populations a clinical trial might
also investigate multiple doses of the same treatment. Dose-finding
trials are routinely performed in Phase II of clinical development and in some cases
Phase III trials also investigate several different doses of the same treatment.
Frequently there might be uncertainty about the underlying dose-response
shape and thus which dose-response model to fit to the data. 
Subgroup identification methods in the context of dose-response trials have recently been introduced in \cite{thom:born:seib:2018} and 
a testing approach, that can be used for dose-response-trials 
with a prespecified subgroup would be useful in practice as well.
In this paper we want to propose such an approach, which is based
on the MCP-Mod methodology (\cite{bret:pinh:bran:2005}).

The MCP-Mod approach combines multiple comparison
procedures and modeling to analyse dose-response trials. MCP-Mod consists of
two parts. In the multiple comparisons procedure (MCP) step a test for a
dose-response signal is performed for several
prespecified candidate dose-response shapes . 
The contrasts for the tests are chosen to achieve
optimal power for the tests. MCP-Mod uses the
joint distribution of the test statistics to 
control the family-wise error rate (FWER) at a specified level, while
accounting for the correlations between test statistics.
In case of a normally distributed outcome the joint distribution of the
tests statistics is known and the correlation structure between the test
statistics depends on the candidate shapes used. The more
general case, which is described in \cite{pinh:born:glim:2014} extends
the framework to non-normal outcomes using the
asymptotic distribution of the test statistics. 
If a dose-response signal is established under any of the
considered shapes, one can continue to the second step, the modeling
part. In the modeling (Mod) step significant models can then be 
fit to the data and be used for dose estimation.
Further details concerning the design and analysis of
clinical trials with MCP-Mod are discussed in
\cite{pinh:born:bret:2006} and \cite{xun:bret:2017}.

The approach we propose in this paper is an extension of the MCP-part of the MCP-Mod
methodology to allow for testing for a dose-response signal in
multiple populations. 
We specifically focus on the situation of a Phase II trial, where there is
uncertainty about the underlying dose-response shape and a
subgroup with a possibly enhanced treatment effect
has been prespecified. The proposed methods adjust
for the multiplicity resulting from tests for different
dose-response shapes and tests in multiple populations, and make use of
the correlation between the tests. We also take into account,
that variance might differ between populations.
While we focus on tests in one subgroup and the full population,
in general the methodology presented here can be used to perform 
multiple comparisons across arbitrarily overlapping populations of
any number, while controlling the FWER. 

The remainder of this paper is structured as follows:
In the following Section we will present the methodology
to perform contrast tests in multiple populations.
In Section 3 we show the results of simulations, which evaluate
the operating characteristics of the approach, focusing on
FWER control and power. In Section 4 
we summarize and  discuss the results.

\section{Multiple contrast tests for multiple populations}
\label{sec:methods}

\subsection{Notation, models and contrast tests}
\label{sec:notmodcontr}
We consider a clinical trial, which has been performed in parallel
groups of patients, which receive different doses $d_1, d_2,..., d_k$
of an experimental treatment. The number of patients in the dose groups are
$n_1,...,n_k$ and $n := \sum\limits_{i=1}^k n_i$, the total number of
patients in the study. We observe responses $\boldsymbol{Y}$, which can
either be related to efficacy or safety of the new treatment. In the following
we distinguish between three populations of interest: the full or overall
population ($F$), which includes all patients, and two prespecified subpopulations of $F$; a
subpopulation of patients, which has been identified to possibly show an
enhanced treatment effect ($S$) and the complement of that subpopulation
($C$), so that $F = S\cup C$. To simplify the discussion of the
methodology we introduce the two index sets, $U := \{F,
S, C\}$ and $V:= \{S, C\}$. In what follows we assume that the
patients can be classified into subgroup and complement without error.

We adopt the general framework from \cite{bret:pinh:bran:2005}, but
additionally introduce the population as an additional component in the
models. Analogously to the original MCP-Mod framework we therefore
consider the model
\begin{equation}
\begin{split}
Y_{ij}^{(P)} = f(d_i, \boldsymbol{\theta})^{(P)} + \epsilon_{ij}^{(P)}; \epsilon_{ij}^{(P)} \sim N(0, \sigma_P^2) \hspace{0.5cm} i.i.d,\\
 i = 1,...,k, j = 1,...,n_i^{(P)}, P\in V
\end{split}
\label{eq:drmod}
\end{equation}
where $\boldsymbol{\theta}$ refers to the vector of model parameters,
$i$ to the dose group and $j$ to the patient within dose group $i$ and
population $P$. $n_i^{(P)}$ denotes the number of patients in population
$P$, which are in dose group $i$, so that $n_i = n_i^{(F)} = n_i^{(S)} + n_i^{(C)}$.
To simplify the discussion we assume that the prevalence of the subgroup 
is deterministic and constant across all dose levels.
We denote the prevalence of $S$ by $\gamma:=\frac{n_1^{(S)}}{n_1^{(F)}} = 
\dots = \frac{n_k^{(S)}}{n_k^{(F)}}$. 
A more general discussion without the assumption of constant prevalences is
included in the Appendix.

Model (\ref{eq:drmod}) is assumed to be the underlying dose-response 
model with regard to dose estimation. For the contrast tests, 
which have the aim of detecting a dose-response relationship, a different model,
\begin{equation}	
\begin{split}
Y_{ij}^{(P)} =\mu_i^{(P)} + \epsilon_{ij}^{(P)}; \epsilon_{ij}^{(P)} \sim N(0, \sigma_P^2) \hspace{0.5cm} i.i.d,\\
 i = 1,...,k, j = 1,...,n_i^{(P)}, P\in V  
\end{split}
\label{eq:drmod2}
\end{equation}
is considered. $\mu_i^{(P)}$ is the mean response for dose group $i$
in population $P$. An estimate for $\mu_i^{(P)}$ is the arithmetic mean
$\bar{Y_i}^{(P)} = \sum\limits_{j =
1}^{n_i^{(P)}}\frac{Y_{ij}^{(P)}}{n_i^{(P)}}$. For population $P$ we
denote the vector of the estimated dose means by $\bar{\boldsymbol{Y}}^
{(P)} = (\bar{Y_1}^{(P)}, \dots, \bar{Y_k}^{(P)}) ^{'}$. Additionally
the pooled variance estimator for $\sigma_P^2$ is 
\begin{equation}
{S ^ {(P)}}^2 = \sum\limits_{i = 1}^k
\sum \limits_{j = 1}^{n_i^{(P)}} (Y_{ij}^{(P)} - \bar{Y_i}^{(P)})^2/
\sum\limits_{i=1}^k (n_i ^ {(P)} - k).
\label{eq:pooledvar}
\end{equation}

We assume uncertainty regarding the underlying shape of the
dose-response relationship and assume, that a set of plausible candidate
models $\{M_1,\dots, M_Z\}$ is prespecified. These models are candidates
to describe the shape of the dose-response curve and produce
mean vectors under the alternative. Usually candidate models are commonly encountered dose-response
models like $E_{max}$, linear, exponential or quadratic shapes. 
These candidate models are used to obtain the contrast
vectors for the tests.
Contrast vectors $\boldsymbol{{c_m}^{'}} =
(c_{m1}, \dots, c_{mk})$ are chosen optimally to guarantee
maximal power for each single contrast test. 
For details on choosing the optimal contrasts see
\cite{bret:pinh:bran:2005}. 

The MCP-Mod
approach then tries to detect a significant dose-response relationship for
at least one of the models. In theory it
is possible to specify a different set of candidate shapes for each 
population. We assume here, that the set of candidate
models is the same across all populations, which is likely
plausible for most real-life situations. With this assumption
and the assumption of a constant prevalence of the subgroups
across all dose groups the optimal contrast vectors are independent of the population, 
in which the test is performed. We discuss the more
general situation of different candidate models across populations
in the Appendix.

For each candidate model $m\in \{M_1,\dots, M_Z\}$ and each population $P\in U$ a contrast
test testing the null hypothesis $H_0^{(P, m)}: \boldsymbol{{c_m}^ {'}
\mu^{(P)}} = 0$ against the alternatives $H_1^{(P, m)}: \boldsymbol{{c_m}^{'} 
\mu^{(P)}} > 0$ is performed. Using this alternative we assume without loss of
generality, that the treatment effect will have a positive sign.
 
In total the number of tests
is $3Z$.  We define the \textit{population null hypothesis (in $P$)} as
\[
H_0^{(P)}: H_0^{(P, M_1)} \cap ... \cap  H_0^{(P, M_Z)},
\]
the intersection of the null hypotheses for all candidate models in population
$P$. The \textit{global null hypothesis} is then the intersection
between the population null hypotheses, in all tested populations.
For example for the full testing strategy, which includes both subgroup
and complement the global null hypothesis is
\[
H_0^{(global)}: H_0^{(F)} \cap H_0^{(S)} \cap H_0^{(C)}
\]
While rejecting the global null is of main interest, the population hypotheses
have to be considered to determine if a dose-response trend has been established 
in all populations or, for example, only in a subgroup.
 
The test statistics are contrast tests of the form
\begin{equation}
T_m^{(P)} = \frac{(\boldsymbol{c_m})^{'}\bar{\boldsymbol{Y}}^{(P)}}{\sqrt{(\hat{\sigma}^{(P)})^2
\sum\limits_{i = 1}^{k} c_{mi}^2/n_i^{(P)}}} \hspace{0.5cm} m\in \{M_1,\dots, M_Z\}, P\in V,
\label{eq:teststats}
\end{equation}
which are performed for each of the candidate models in each population.
$(\hat{\sigma}^{(P)})^2$ is an estimator of the variance in population
$P$. There are
several estimators that could be considered. 
Which estimator to use depends on the assumptions made regarding possible 
heteroscedasticity between subgroup and complement and has 
an effect on the critical values used for the tests as we will
discuss in the following.

If the variances in subgroup and complement are assumed to be the
same, e.g. we assume homoscedasticity between populations,
we can drop the population indices for the variances. Then
$\sigma_S^2 = \sigma_C^2 = \sigma^2$. We can then also 
drop the population index from the variance estimator 
in (\ref{eq:teststats}) and use the same estimator for all 
tests. A pooled variances estimator (pooled over dose levels
and over subgroup and complement) is the natural choice
in this context, since model (\ref{eq:drmod2}) assumes that
the means are different both across dose levels, and subgroup and
complement. The pooled variance estimator is then
\begin{equation}
\hat{\sigma}^2 =
\sum\limits_{P\in V}\sum\limits_{i = 1}^k \sum \limits_{j = 1}^{n_i^{(P)}}
(Y_{ij}^{(P)} - \bar{Y_i}^{(P)})^2/ \sum\limits_{P\in V}(\sum\limits_{i=1}^k
n_i ^ {(P)} - k).
\label{eq:pooledvar}
\end{equation}

The assumption of homogeneous variances in all populations may not
always be justified. For example patients in a subgroup might be assumed
to be more homogeneous than patients in the full population and thus
the subgroup will be assumed to have smaller variances.
If we assume the variances in subgroup and complement to be unequal,
e.g. we assume heteroscedasticity between subgroup and complement
($\sigma_S^2 \neq \sigma_C^2$)
the pooled variance estimator will generally be biased and different
estimators should be used. 

When heteroscedasticity is assumed, it is more appropriate
to estimate the variance separately in subgroup and complement. 
In each population a variance estimate, which is pooled 
over the dose levels as in (\ref{eq:pooledvar}) can then be used.
Then $(\hat{\sigma}^{(S)})^2 = (S ^ {(S)})^2$ and
$(\hat{\sigma}^{(C)})^2 = (S ^ {(C)})^2$.

Under the global null hypothesis there is no difference between
contrasts in subgroup and complement (both are assumed to be zero).
Under these assumptions a weighted average of the two population variance
estimates is a reasonable estimator for the variance in the
full population. For the tests in the full population under assumed
heteroscedasticity we would therefore
use the variance estimator
$(\hat{\sigma}^{(F)})^2 = \frac{1}{n}\cdot
[\sum\limits_{i=1}^k n_i ^ {(S)}(\hat{\sigma}^{(S)})^2 +
\sum\limits_{i=1}^k n_i ^ {(C)}(\hat{\sigma}^{(C)})^2)]$.

\subsection{FWER control}
\label{sec:fwer}
Performing multiple tests at once, as in the approach we
described above, leads to multiplicity issues. 
In the classic MCP-Mod approach the family-wise error rate (FWER)
across all tests is controlled, e.g. the probability to reject at least
one true null hypothesis is below a specified level $\alpha$.
The MCP-Mod approach makes use of the joint distribution
of the test statistics to obtain critical values or adjusted $p$-values. 
In the situation we consider here, in addition to the multiplicity
stemming from testing different candidate shapes, there is
also multiplicity stemming from tests in several populations. 
When MCP-Mod is performed in only one population,
the joint distribution is known to be multivariate $t$ under
the assumptions of model (\ref{eq:drmod2}).
When considering multiple populations the joint distribution
of the test statistics is only multivariate $t$, if the variance is assumed
to be equal across all populations.

In the following discussion we assume that tests are performed
in the full population, the subgroup and the complement, since
this is the most complete testing strategy and the distributions
for other strategies can easily be derived from this.

We will first consider the homoscedastic case, where variances are assumed
to be equal in subgroup and complement 
($\sigma_S^2 = \sigma_C^2 = \sigma^2$).
Then it is appropriate to use the pooled variance estimate (\ref{eq:pooledvar})
in all populations and the vector
of test statistics $\boldsymbol{T ^ {'}} = (T_{M_1}^{(F)},
\dots, T_{M_Z}^{(F)}, T_{M_1}^{(S)}, \dots, T_{M_Z}^{(S)}, 
T_{M_1}^{(C)}, \dots, T_{M_Z}^{(C)})$ is distributed as
$MVT_{3Z}(\nu, \mathbf{0}, \boldsymbol{R})$ with $\nu =
\sum\limits_{P\in V}(\sum\limits_{i=1}^k n_i ^ {(P)} - k)$ degrees of
freedom, mean vector $\mathbf{0}$ and correlation matrix $\boldsymbol{R}$. 
The correlation matrix has the form
\[
\boldsymbol{R} =
\begin{bmatrix}
	\boldsymbol{R_{FF}} & \boldsymbol{R_{FS}} & \boldsymbol{R_{FC}} \\
        \boldsymbol{R_{FS}^{'}} & \boldsymbol{R_{SS}} & \boldsymbol{R_{SC}} \\
	\boldsymbol{R_{FC}^{'}} & \boldsymbol{R_{SC}^{'}} & \boldsymbol{R_{CC}} \\
\end{bmatrix}, 
\]
where the entries of submatrix $\boldsymbol{R_{PQ}}\in\mathbb{R}^{Z \times Z}$ 
for $P, Q \in U$ are given as
\[
(\rho_{ij})^{(PQ)} = \frac{\sum\limits_{l =
1}^Kc_{il}c_{jl}\cdot n_l^{(P \cap
Q)}/(n_l^{(P)}n_l^{(Q)})}{\sqrt{\sum\limits_{l =
1}^K{c_{il}}^2/n_l^{(P)}\sum\limits_{l =
1}^K{c_{jl}}^2/n_l^{(Q)}}}, \hspace{0.5cm} i,j=1,\dots, Z.
\]
Based on the above formula it is easily visible, that 
$\boldsymbol{R_{SC}} = \boldsymbol{0}$, since subgroup and complement
are distinct. The other submatrices are easily obtainable as well, for example
\[
(\rho_{ij})^{(FS)} = \sqrt{\gamma}\cdot \frac{\sum\limits_{l =
1}^kc_{il}c_{jl}
/n_l^{(F)}}{\sqrt{\sum\limits_{l =
1}^k{c_{il}}^2/n_l^{(F)}\sum\limits_{l =
1}^k{c_{jl}}^2/n_l^{(F)}}}, \hspace{0.5cm} i,j=1,\dots, Z
\]
follows for the entries of the matrix $\boldsymbol{R_{FS}}$.

The joint distribution for  performing
tests only in the full population and in the subgroup without a separate
test in the complement can easily be derived by dropping
the irrelevant test statistics from the vector $\boldsymbol{T ^ {'}}$
and removing the irrelevant submatrices (in this case the last row and column)
 from $\boldsymbol{R}$. Using the quantiles of the joint distribution as critical
values for the tests will then guarantee strong control of the FWER
over the population null hypotheses.

If variances cannot assumed to be equal in subgroup and 
complement, the situation becomes more challenging. 
In this case the variance estimator in the denominator 
of the test statistics (\ref{eq:teststats}) is population-dependent and the degrees
of freedoms of the $t$-distributions are no longer the same in all 
populations. Therefore the joint distribution is not multivariate
$t$ and in fact does not seem to belong to any standard probability distribution.
The terms in the correlation matrix $\boldsymbol{R}$ also become
more complex, since the variance 
terms no longer cancel out for all of the submatrices. In general
the correlation submatrices then have entries 

\begin{gather*}
\scalebox{1.1}{$
(\rho_{ij})^{(PQ)} = \frac{\sum\limits_{l =
1}^Kc_{il}c_{jl}\cdot (n_l^{(P \cap
Q\cap S)}\sigma_S^2 + n_l^{(P \cap
Q\cap C)}\sigma_C^2)/(n_l^{(P)}n_l^{(Q)})}{\sqrt{\sum\limits_{l =
1}^K\{{c_{il}}^2\cdot (n_l^{(P \cap S)}\sigma_S^2 + n_l^{(P \cap C)}\sigma_C^2)/{n_l^{(P)}}^2\}
\sum\limits_{l =1}^K\{{c_{jl}}^2\cdot (n_l^{(Q \cap S)}\sigma_S^2 + n_l^{(Q \cap C)}\sigma_C^2)/{n_l^{(Q)}}^2}\}}, \hspace{0.5cm}$}\\
 i,j=1,\dots, Z; P, Q \in U.
\end{gather*}

For example for the entries of submatrix $\boldsymbol{R_{FS}}$
we get
\[
(\rho_{ij})^{(FS)} = \frac{\sigma_S\sum\limits_{l =
1}^Kc_{il}c_{jl}/n_l^{(F)}}{\sqrt{[\gamma\sigma_S^2 + (1-\gamma)\sigma_C^2]
\sum\limits_{l =1}^K{c_{il}}^2 /n_l^{(F)}
\sum\limits_{l =1}^K{c_{jl}}^2/n_l^{(S)}}}, \hspace{0.5cm}\\\
 i,j=1,\dots, Z.
\]
The variances $\sigma_S^2$ and $\sigma_C^2$ are usually unknown. 
To obtain the correlation matrix, the estimates for $\sigma_S^2$ and $\sigma_C^2$ 
can be plugged into the formulas above.

There are several reasonable options to approximate the joint
distribution of the tests statistics under heteroscedasticity and
obtain multiplicity-adjusted $p$-values.
A simple solution is to assume that
the variances are known and use the multivariate normal distribution to
obtain $p$-values and critical values. Since the multivariate $t$
distribution approaches a multivariate normal distribution as $n
\rightarrow \infty$ this approximation will work well
for reasonably large sample sizes. 

For small sample sizes a conservative option
is to use the smallest degrees of freedom across all populations, 
$\nu_{min} := \min\limits_{P\in V} \nu^{(P)}$, 
where $\nu^{(P)} := \sum\limits_{i=1}^k n_i ^ {(P)} - k$
are the population specific degrees of freedom. In the considered
setting this is equivalent to using the degrees of freedom from
the population with the smallest number of patients.
Then a $MVT_{3Z}(\nu_{min}, \mathbf{0}, \boldsymbol{R})$ -
distribution can be used as an approximate joint distribution.

A less conservative approximation, which is proposed in \cite{hasl:2014}
is to use a \textit{multiple degrees of freedom} approximation, which translates to
using a different multivariate $t$-distribution with the correct degrees
of freedom for each population to obtain critical values. 
E.g. for test statistics from population
$P\in U$, $T_{M_1}^{(P)}, \dots, T_{M_Z}^{(P)}$, the critical values are obtained
from a $MVT_{3Z}(\nu^{(P)}, \mathbf{0}, \boldsymbol{R})$ -
distribution. A similar approach has also been discussed 
by \cite{graf:wass:frie:2018} for standard (non-contrast)
$t$-tests in subgroups, while additionally allowing for non-fixed
sample sizes in subgroups.

Table \ref{tab:methods} summarizes all discussed methods both
under the assumption of homoscedasticity and heteroscedasticity.
\begin{table}
\centering
\caption{Overview over considered MCPs with the variance estimates used in the test statistics, the distribution used to obtain critical values and the degrees of freedom
for the multivariate $t$ distributions. Here $\nu^{(P)} := \sum\limits_{i=1}^k n_i ^ {(P)} - k$, the degrees of freedom for population $P$.}
\scriptsize
  \begin{tabular}{l|l|l|l|l}
    Method & Description & Variance estimation & Distribution & Degrees of freedom\\\hline
    SP & \parbox[t]{4cm} {Standard MCP in \\ single (full) population} & pooled & multivariate $t$ &  $\sum\limits_{i=1}^k n_i ^ {(F)} - k$ \\
    MP-Pooled &\parbox[t]{4cm} { Multi-population MCP \\assuming homoscedasticity \vspace{0.15cm}} & pooled & multivariate $t$ &  $\sum\limits_{P\in V}\sum\limits_{i=1}^k \nu^{(P)}$ \\
    MP-MinDF &\parbox[t]{4cm} { Multi-population MCP \\assuming heteroscedasticity \vspace{0.15cm}} & population-specific & \parbox[t]{3cm} {multivariate $t$ \\(approx.)} & $\min\limits_{P\in V} \nu^{(P)}$ \\
    MP-MultDF & \parbox[t]{4cm} {Multi-population MCP \\assuming heteroscedasticity \vspace{0.15cm}} & population-specific & \parbox[t]{3cm} {multivariate $t$ \\(approx.)} & \parbox[t]{4cm} {$\nu^{(P)}$ (different \\ for each population)}\\
    MP-Normal & \parbox[t]{4cm} {Multi-population MCP \\assuming heteroscedasticity \vspace{0.15cm}} & population-specific & \parbox[t]{3cm} {multivariate normal \\(approx.)} & -\\
  \end{tabular}

\label{tab:methods}
\end{table}

\section{Simulation study}
\label{sec:sims}

In this section we discuss results of a simulation study
to evaluate the properties of the multi-population
testing approaches. The simulation setup is similar
to the one used in \cite{bret:pinh:bran:2005}. 
The simulations are divided into two
main parts. In the first part (Section \ref{sec:sim_homvar})
we assume homoscedasticity across
populations and evaluate the power of the proposed multi-
population testing approaches compared to the standard single population MCP, which
only considers the full population. The second part (Section \ref{sec:sim_hetvar}) 
deals with the scenario of heteroscedastic populations. In addition
to the comparison of multi-population  MCP to single population
MCP the main focus of this Section is to evaluate the different considered
procedures, which are based on approximations of the joint
distribution, with regard to FWER control and power. 

\subsection{Simulation setup}
\label{sec:simsetup}
We simulate clinical trials with parallel groups,
which are balanced over the dose levels. In our simulated trial patients
are randomized to one of five different dose groups with dose levels
0, 0.05, 0.2, 0.6, 1.  We assume that there is knowledge about a subgroup,
which possibly shows an enhanced treatment effect and could have
been identified from an earlier trial or based on the mode of action of
the drug. We also make the simplifying assumption, that for each
dose group the proportion of patients in the subgroup is equal. 
Together with the assumption of a balanced trial design, this
means, that  $n_1^{(S)} = \dots = n_k^{(S)}$.

We generate the observed data for the simulated trials from models
of the form
\begin{equation}
\begin{split}
	Y_{ij}^{(P)} = f(d_i, \boldsymbol{\theta})^{(P)} + \epsilon_{ij}^{(P)}; \epsilon_{ij}^{(P)} \sim N(0, \sigma_P^2) \hspace{0.5cm} i.i.d,\\
 i = 1,...,k, j = 1,...,n_i^{(P)}, P \in V
\end{split}
\label{eq:drmod_sims}
\end{equation}
and use common dose-response shapes for $f$. An overview over the used
dose-response shapes is given in Table \ref{tab:candshap}. We also include a constant
model to generate data under the global null hypothesis. 
\begin{table}
\centering
\caption{Data generating dose-response models and 
corresponding standardized versions, which are used as candidate shapes
to determine optimal contrast tests. Parameters specified in the last column are
fixed for all simulated trials and are used as guesstimates to derive
optimal contrasts.}
  \begin{tabular}{l|l|l|l}
    Model & $f(d, \boldsymbol{\theta})$ & standardized model & fixed parameters\\\hline
    Constant & 0.2 & - & -\\
    Emax & $\theta_0 + \frac{\theta_1d}{(\theta_2 + d)}$ & $\frac{d}{(\theta_2 + d)}$ & $\theta_0 = 0.2$; $\theta_2 = 0.2$\\
    Linear & $\theta_0 + \theta_1d$ & $d$ & $\theta_0 = 0.2$\\
    Exponential & $\theta_0 + \theta_1 \cdot exp[\frac{d}{\theta_2} - 1]$ &  $exp[\frac{d}{\theta_2} - 1]$ & $\theta_2 = 0.29$\\
    Logistic & $\theta_0 + \frac{\theta_1}{1 + exp[(\theta_2 - d)/\theta_3}$ & $\frac{1}{1 + exp[(\theta_2 - d)/\theta_3}$ & $\theta_2 = 0.4$; $\theta_3 = 0.091$\\
    Quadratic & $\theta_0 + \theta_1d + \theta_2d^2$ & $d + \frac{\theta_2}{\lvert \theta_1\rvert}d^2$ & $\frac{\theta_1}{\theta_2} = 1.171$\\
  \end{tabular}

\label{tab:candshap}
\end{table}
The subgroup we simulate has prevalences $\gamma$ of 
either 0.25, 0.5 or 0.75.
We assume that the underlying shape is the same for subgroup
and complement. However the 'slope' parameters of the model, which are not fixed
(see Table \ref{tab:candshap}) can vary between populations.
This allows us to simulate scenarios with different treatment effects
between subgroup and complement.

We consider three scenarios for how the treatment effect is distributed 
between subgroup and complement. 
In the first scenario there is no enhanced
effect in the subgroup, e.g. the treatment effect is identical in the
subgroup and the complement (\textit{same}-scenario). 
In the second scenario the treatment effect
in the subgroup is double the size of the effect in the complement
 (\textit{double}-scenario). 
Finally for the third and last scenario there is only a treatment 
effect in the subgroup and the treatment effect in the complement is zero
(\textit{only}-scenario). The three
different scenarios represent situations, in which there
is in reality no subgroup and possible treatment effects
are constant across the full population (\textit{same}) , 
or in which a treatment effect in the
full population is driven mostly (\textit{double}) or completely (\textit{only}) by a large effect
in the subgroup. 

We choose the (non-fixed) parameters for the non-constant models,
so that the treatment effect at the highest dose is 0.6, which
leads to a power of 80\% (for a pairwise comparison)
for a group size of 75.
For the \textit{same}-scenario 0.6 is the treatment effect in the 
full population, for the \textit{only}- and \textit{double}-scenarios 
it is the treatment effect in the subgroup. This leads to scenarios,
where the treatment effect is relatively large in the subgroup but possibly
hard to detect (depending on $\gamma$) in the full population.

For the multiple comparison procedure we then perform
tests for all non-constant shapes in Table \ref{tab:candshap} 
in all populations we consider,
irrespective of which of the shapes was used for data-generation.
For the multi-population MCP approaches we compare two different
testing strategies. These testing strategies differ with regards
to the populations in which tests are performed. Tests are
performed either in the full population and the subgroup ($F+S$) or in full population, 
subgroup and the complement ($F+S+C$). The first  strategy would be used in 
a situation, where there is information about a possibly existing
subgroup, which could potentially have an enhanced treatment effect.
The second strategy could be applied in a similar situation, but could
additionally be used to confirm that the treatment effect in the
full population is not only driven by a subgroup with an enhanced effect.
If the test could also be rejected in the complement then this would confirm 
that there is consistency across the whole population with regards to the
treatment effects.

\subsection{Homoscedastic scenarios}
\label{sec:sim_homvar}
We will first discuss the simulation results for equal variances
in subgroup and in the complement. For these
simulations we fix the standard deviations at 
$\sigma_S = \sigma_C = 1.478$. 
The results discussed in this Section are based on 5000 simulated trials.
In the article only selected results are shown. Complete simulation
results for all data-generating models are included in the Supporting
Information.

In the homoscedastic case  the exact joint
distributions of the test statistics can be used to obtain critical values
or adjusted $p$-values for the individual tests as described in 
Section \ref{sec:fwer}.
Since the exact distribution of the tests statistics is known, FWER is controlled
at the nominal $5\%$-level for all testing methods, 
when data are generated under the global null hypothesis, from
a constant model. This is confirmed in the simulation results, when data
are generated from a constant model (shown in Supporting Information).

\begin{figure}[h!]
\centering
\includegraphics[width=\textwidth]{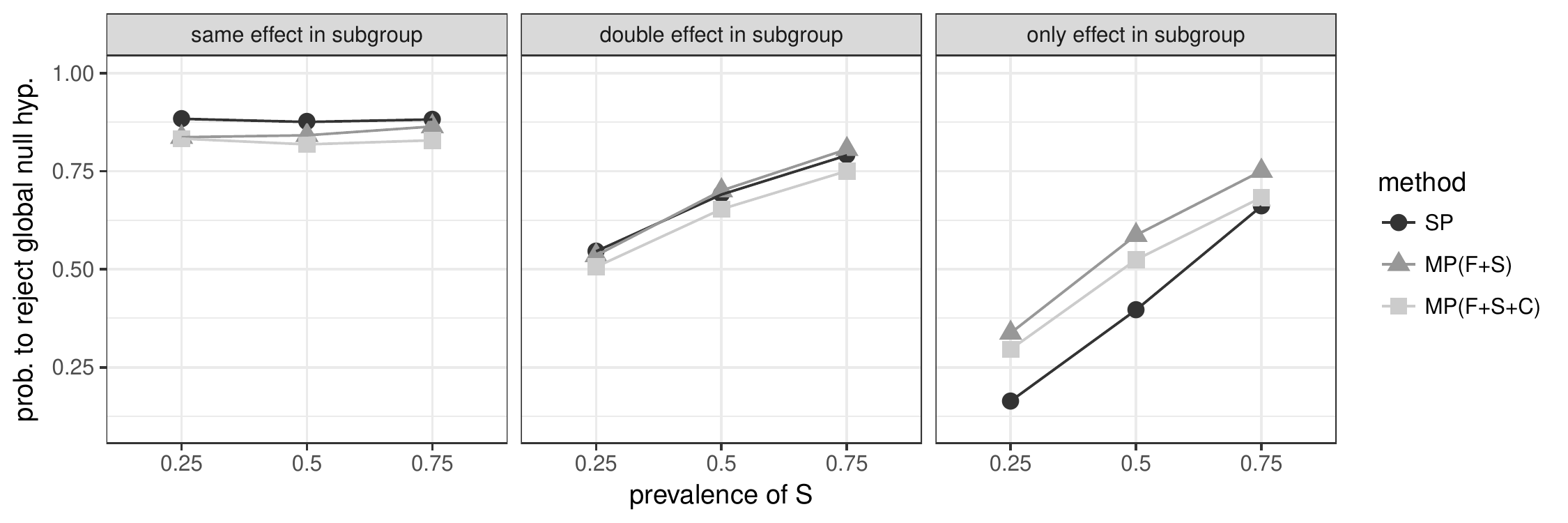}
\caption{Probability to reject the global null hypothesis for single population (SP) and
multi-population (MP) testing methods. Data
are generated from an $E_{max}$ model under homoscedasticity.}
\label{fig:poa_hom_emax}
\end{figure}

When we simulate data under the alternative of
an existing dose-response trend, the MP
testing approach with tests in full population and subgroup increases 
the chance of rejecting the global null hypothesis, when
the treatment effect only exists in the subgroup or is doubled in the subgroup.
This is visualized for data generated from an $E_{max}$ model 
in Figure \ref{fig:poa_hom_emax}. The MP approach with tests
in all populations (F + S + C) also shows more rejections than
SP in the \textit{only}-scenario.
In the \textit{same}-scenario SP
rejects most often, but the loss in power for the MP methods is
relatively small, especially when the subgroup is large.

For the MP methods it might
additionally be of interest, for which population(s) a 
significant dose-response relationship is declared. For example a
significant result only in the subgroup, but not in the full population
or the complement might lead to very different conclusions than significant results in all populations. 
Population specific rejection results for data generated from $E_{max}$ models are summarized
in Table \ref{tab:poa_where_hom_emax}.
The results show that when the treatment effect only exists in the subgroup,
the null hypothesis in the subgroup is rejected up to $24\%$ more often
than the null hypothesis in the full population. The null hypothesis for the complement,
which is true in the \textit{only}-scenario is only rejected in $2\%$ of
simulations.
When the treatment effect in the subgroup is doubled over the complement
rejections generally occur more often in the full population than in the subgroup.
The difference decreases with increasing subgroup prevalence $\gamma$.
Table \ref{tab:poa_where_hom_emax} also includes the results for the MP-MultDF
method, which would also allow for possible heteroscedasticity. The results are essentially
the same as for MP-Pooled, there does not seem to be a big penalty for allowing
for heteroscedasticity in homoscedastic scenarios.
\begin{table}[ht]
\centering
\caption{Probability to reject the global null hypothesis and 
population null hypotheses for single population (SP) and
multi-population (MP) testing methods. Data
are generated from an $E_{max}$ model under homoscedasticity.}
\small
\begin{tabular}{ll|lllllllll}
& & \multicolumn{8}{c}{Scenario and Prevalence}\\
& & \textit{same} & \textit{same} & \textit{same} & \textit{double} & \textit{double} & \textit{double} & \textit{only} & \textit{only} & \textit{only} \\ 
Method & Hypothesis & 0.25 & 0.5 & 0.75 & 0.25 & 0.5 & 0.75 & 0.25 & 0.5 & 0.75 \\ 
  \hline
SP & global & 0.88 & 0.88 & 0.88 & 0.55 & 0.69 & 0.79 & 0.16 & 0.40 & 0.66 \\ 
  MP-Pooled(F+S) & global & 0.84 & 0.84 & 0.86 & 0.54 & 0.70 & 0.81 & 0.34 & 0.59 & 0.75 \\ 
   & F & 0.82 & 0.82 & 0.85 & 0.44 & 0.60 & 0.74 & 0.11 & 0.31 & 0.60 \\ 
   & S & 0.31 & 0.55 & 0.73 & 0.30 & 0.54 & 0.73 & 0.30 & 0.55 & 0.72 \\ 
  MP-Pooled(F+S+C) & global & 0.83 & 0.82 & 0.83 & 0.51 & 0.65 & 0.75 & 0.30 & 0.52 & 0.68 \\ 
   & F & 0.79 & 0.77 & 0.79 & 0.40 & 0.55 & 0.67 & 0.09 & 0.26 & 0.53 \\ 
   & S & 0.28 & 0.47 & 0.66 & 0.27 & 0.48 & 0.67 & 0.26 & 0.49 & 0.65 \\ 
   & C & 0.65 & 0.47 & 0.27 & 0.21 & 0.15 & 0.08 & 0.02 & 0.02 & 0.02 \\  
 MP-MultDF(F+S) & global & 0.83 & 0.84 & 0.87 & 0.53 & 0.69 & 0.80  & 0.35 & 0.59 & 0.76 \\ 
   & F & 0.81 & 0.82 & 0.85 & 0.42 & 0.59 & 0.74  & 0.12 & 0.32 & 0.60\\ 
   & S & 0.28 & 0.53 & 0.72 & 0.30 & 0.54 & 0.73  & 0.30 & 0.55 & 0.72 \\ 
  MP-MultDF(F+S+C) & global & 0.83 & 0.82 & 0.83 & 0.49 & 0.64 & 0.75 & 0.31 & 0.53 & 0.69 \\ 
   & F & 0.78 & 0.77 & 0.79 & 0.38 & 0.53 & 0.67  & 0.10 & 0.27 & 0.52 \\ 
   & S & 0.25 & 0.47 & 0.65  & 0.26 & 0.47 & 0.66 & 0.27 & 0.48 & 0.64 \\ 
   & C & 0.65 & 0.47 & 0.26  & 0.20 & 0.14 & 0.08 & 0.02 & 0.02 & 0.02 \\ 
  
\end{tabular}
\label{tab:poa_where_hom_emax}
\end{table}

One of the main features of the discussed methods is the possibility to test for
several possible dose-response shapes. Hence it is also of interest to compare the model selection
performance of the MP testing approaches compared to the SP-MCP.
Table \ref{tab:modsec_hom} in the appendix shows the probability of choosing the
correct model for the different methods. The probability of choosing
the correct model based on the lowest significant $p$-value
is almost identical between SP and MP approaches 
and therefore testing in multiple populations does not seem to reduce the model
selection performance.

\subsection{Heteroscedastic scenarios}
\label{sec:sim_hetvar}
In this part we will show simulation results, when testing in multiple
populations with different variances. Here we use $\sigma_S = 1.03$ and
$\sigma_C = 1.926$ for the standard deviations in 
subgroup and complement. The combined standard deviation in 
the full population thus remains the same as before at 1.478,
as long as subgroup and complement have equal prevalence ($\gamma = 0.5$).

As discussed in Section \ref{sec:fwer} the joint
distribution of the contrast test statistics does not follow a standard
distribution, when variances are heterogeneous 
and instead approximations to the true joint distribution
are necessary to obtain critical values. In these simulations
we will therefore first compare the different possible approximations we discuss
in Section \ref{sec:fwer}. We focus on FWER control and
power for the considered approximations. For these
simulations we increased the number of simulations from
5000 to 25000, to ensure that possible differences between
the methods are distinguishable from simulation error.

The approaches we consider for approximating the joint
distribution are MP-MinDF, MP-MultDF and MP-Normal (see Table
\ref{tab:methods}).
All of these approaches use separate variance estimates for the tests 
in subgroup and complement.
To evaluate the general necessity of using population-specific variance estimates
we also include test statistics using the pooled variance estimate (MP-Pooled), which
assumes, that variances are the same in subgroup and complement.

\begin{figure}[h!]
\centering
\includegraphics[width=\textwidth]{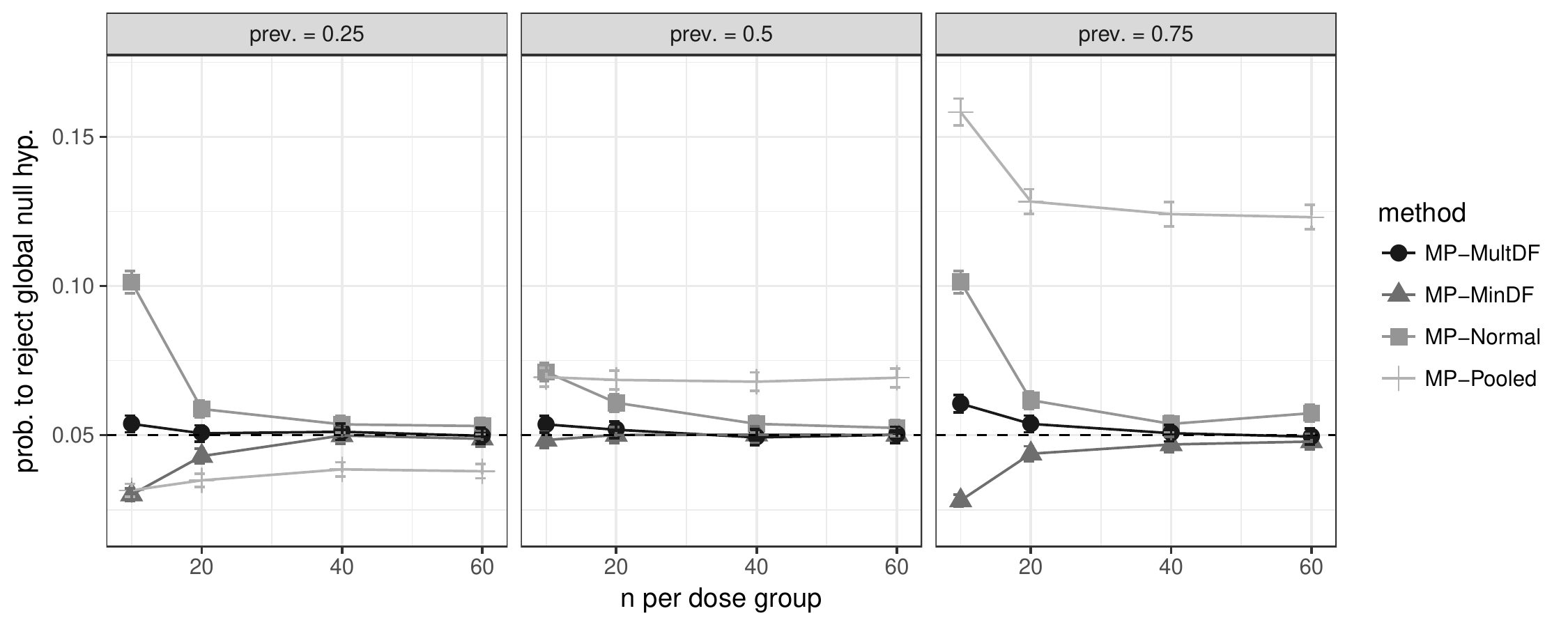}
\caption{Probability to reject the global null hypothesis for different
multi-population (MP) testing methods. Data
are generated from a constant model under heteroscedasticity.
Tests were performed in full population, subgroup and complement.
Bars show the $95\%$ - confidence intervals (accounting for simulation error).}
\label{fig:poa_het}
\end{figure}

Figure \ref{fig:poa_het} shows the probability of rejecting the global null
hypothesis when the global null is true, e.g. data are generated from a constant
model. MP-Pooled, which ignores heteroscedasticity,
generally does not control the FWER at the nominal level, since the variance in the
complement is underestimated, when
the subgroup is large. For a subgroup with prevalence of 0.75 the FWER is generally
above 0.1 with the pooled variance estimate. 
Of the three approximations, which use population-specific variance estimates,
MP-Normal is as expected liberal
and leads to a large inflation of the FWER for small sample sizes.
For large sample sizes MP-Normal approximates the true joint
distribution better and the FWER is close to the nominal level.
Of the two remaining approaches, which use approximate multivariate $t$ distributions,
the MP-MinDF approach controls the FWER rate at the
nominal level for all sample sizes and prevalences.
MP-MultDF shows similar error
rates but shows FWERs slightly higher than 0.05 for very small sample
sizes.

\begin{figure}[h!]
\centering
\includegraphics[width=\textwidth]{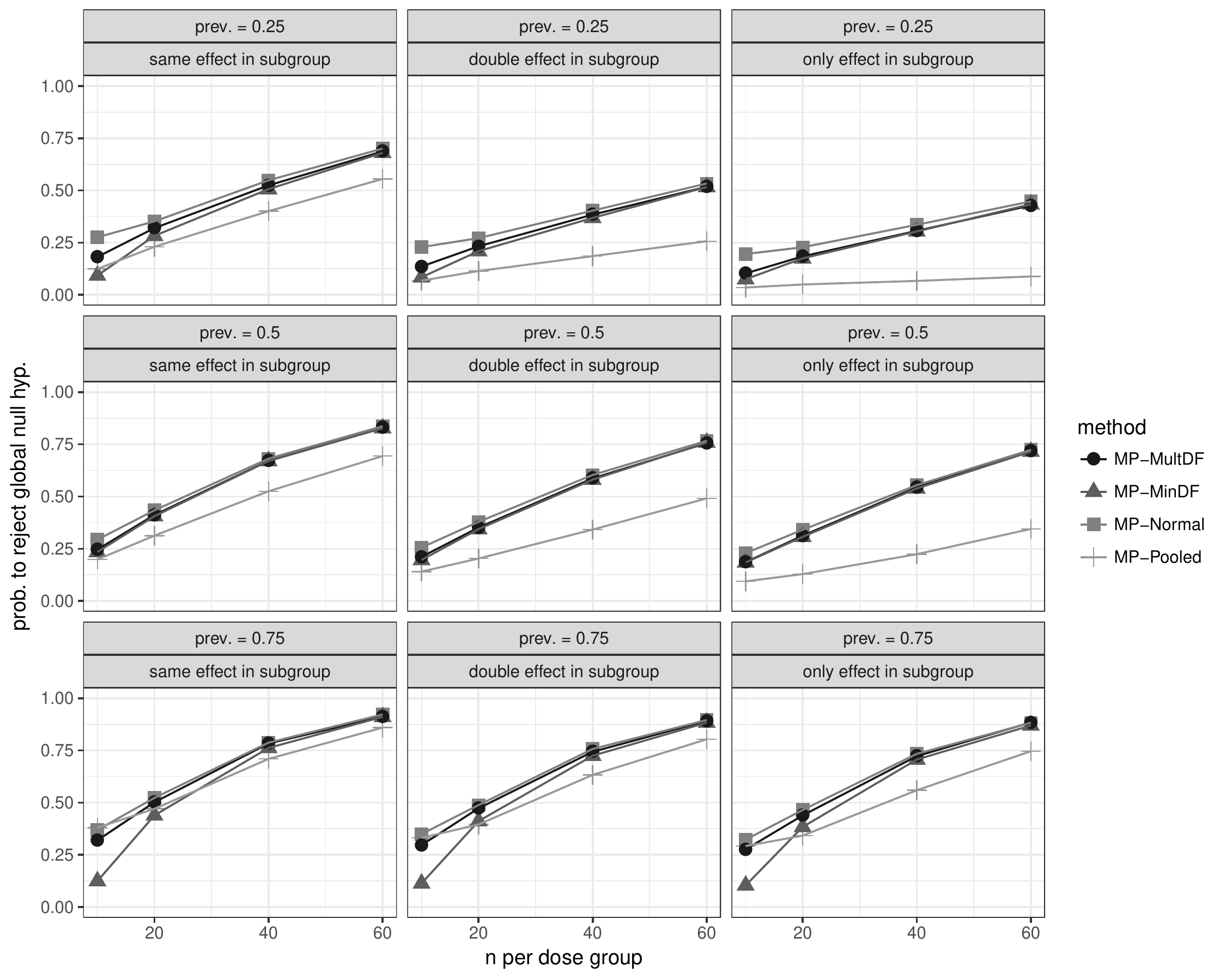}
\caption{Probability to reject the global null hypothesis for different
multi-population (MP) testing methods. Data
are generated from an $E_{max}$ model under heteroscedasticity.
Tests were performed in full population, subgroup and complement.}
\label{fig:poa_het_emax}
\end{figure}

\begin{figure}[h!]
\centering
\includegraphics[width=\textwidth]{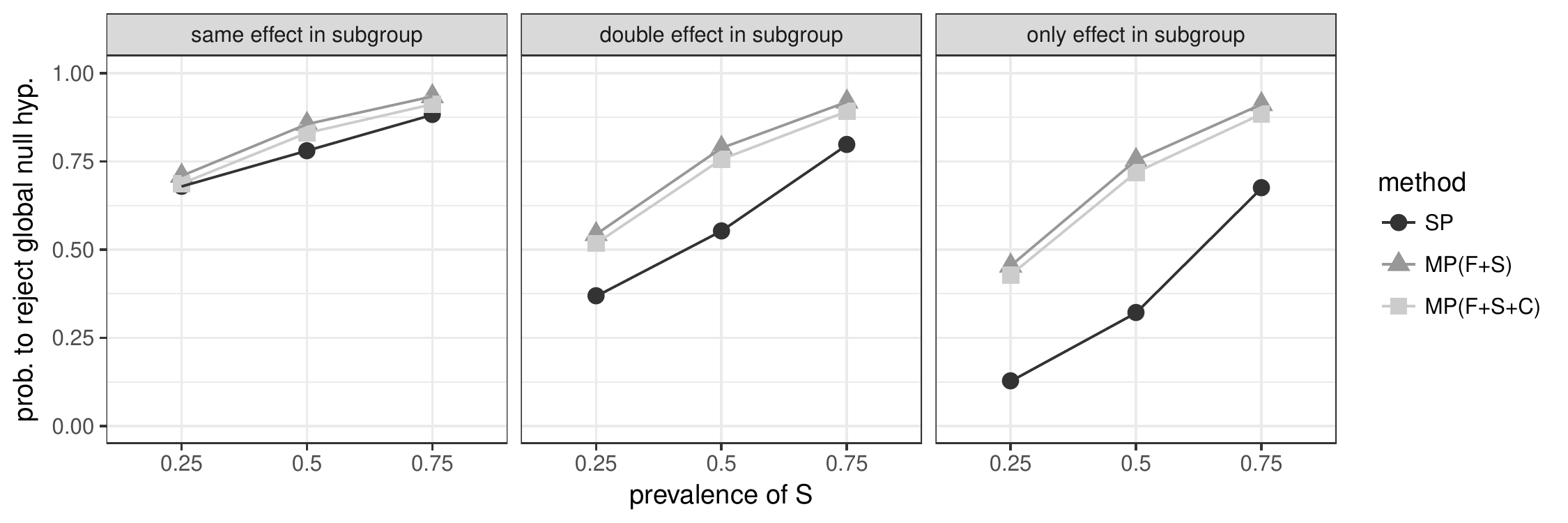}
\caption{Probability to reject the global null hypothesis for single population (SP) and
multi-population (MP) testing methods. Data
are generated from an $E_{max}$ model under heteroscedasticity.
MP-MultDF is used to approximate the joint distribution for
MP testing methods.}
\label{fig:poa_het_emax2}
\end{figure}

Figure \ref{fig:poa_het_emax} visualizes the differences with regards to power,
e.g. the frequency of rejections of the global null hypothesis
for data generated from an $E_{max}$ model for different scenarios and
subgroup prevalences. With pooled variance
estimation the power is generally reduced compared to the other methods. 
The approaches using population-specific variances estimates
show fairly similar power, 
with MP-Normal being more liberal for small sample sizes 
and therefore rejecting more often. The MP-MultDF
approach always leads to slightly more rejections than the 
MP-MinDF approach. The conclusions for the other data-generating
dose-response models are similar, detailed results for these models
can be found in the Supporting Information.
Finally it is of interest to compare the multi-population testing strategies 
to the single population MCP in the heteroscedastic setting
as well. We focus here on the MP-MultDF approach, since it gives FWER 
control roughly at the nominal level, while not sacrificing too much power.
Figure \ref{fig:poa_het_emax2} compares the methods
for a sample size of 60 per group. Comparing Figure \ref{fig:poa_het_emax2}
to Figure \ref{fig:poa_hom_emax} the differences are now even larger
between SP and MP methods. Since SP does not
account for the different variances in subgroup and complement, the MP
methods give better power even when there is no increased treatment effect
in the subgroup and the treatment effect is the same for all patients.

\section{Discussion}
\label{sec:disc}
In this paper we discussed an extension of the MCP part 
of the MCP-Mod 
methodology to allow for testing 
in multiple populations. 
We focused on the situation
of a dose-response trial, where a subgroup with a suspected
enhanced treatment effect has been prespecified. Our method
allows for simultaneous testing in
full population, subgroup and possibly complement under model
uncertainty, while
controlling the FWER across all candidate models and populations.
Our simulation results
in Section \ref{sec:sim_homvar} and \ref{sec:sim_hetvar} 
show, that an increase in power of rejecting the global null hypothesis can 
be achieved, when using the multi-population
approach. The results suggest, that the prevalence of the subgroup 
and the size of the treatment effect in the subgroup relative to
the complement are important factors to determine if a multi-population approach
is worth considering. Additional plots, which visualize the effects of these factors
on the power in more detail are included in the Supporting Information.

Instead of considering the power, one can also use sample
size calculations to illustrate the potential benefits of
a multi-population testing approach.
Table \ref{tab:sampsize} shows the sample
sizes needed to achieve $80\%$ power for rejecting
the global null hypothesis under the scenarios
considered for the simulation study in Section
\ref{sec:sims}. For situations, where the subgroup
has a strongly enhanced effect over the rest of the population,
the sample sizes for the trial can be significantly reduced
when using a multi-population approach instead of 
a single-population approach. 
\begin{table}[ht]
\centering
\caption{Sample sizes (per dose) required to achieve $80\%$ power for
rejecting the global null hypothesis for single population (SP) and
multi-population (MP) testing methods. Required sample sizes are calculated
under the assumption of homoscedasticity across populations.}
\begin{tabular}{l|lllllllll}
& \multicolumn{8}{c}{Scenario and Prevalence}\\
& \textit{same} & \textit{same} & \textit{same} & \textit{double} & \textit{double} & \textit{double} & \textit{only} & \textit{only} & \textit{only} \\  
Method & 0.25 & 0.5 & 0.75 & 0.25 & 0.5 & 0.75 & 0.25 & 0.5 & 0.75 \\ 
  \hline
SP & 59 & 60 & 59  & 151 & 105 & 77 & 932 & 235 & 105\\ 
MP-Pooled(F+S) & 70 & 67 & 63 & 151 & 101 & 75  & 275 & 131 & 84 \\ 
MP-Pooled(F+S+C) & 71 & 70 & 70  & 163 & 112 & 87 & 299 & 147 & 97 \\ 
\end{tabular}
\label{tab:sampsize}
\end{table}

In the heteroscedastic case, the simulation
results discussed
in \ref{sec:sim_hetvar} show, that the multiple degrees of freedom
approach seems to result in a good approximation to the 
joint distribution of the test statistics. The FWER-inflation 
is minimal compared to the
multivariate normal approximation, which is only
suitable for large sample sizes. Additionally, the multiple
degrees of freedom approach sacrifices less power than the 
conservative minimum degrees of freedom approach.

In the introduction we stated three different outcomes for a
trial in multiple populations:
(i) the drug is shown to be efficacious for all patients, 
(ii) the drug is shown to be efficacious only in a subgroup or
(iii)  the drug is shown to be efficacious for all patients but with
an increased efficacy for patients in a subgroup.
The methodology we present here can be used to help with
decision-making in this context. We would advise
to use our methodology together with estimates of
treatment effects in the different populations to decide
which of the three outcomes is achieved.
\cite{mill:dmit:rube:2012} propose a decision framework 
for testing in a subgroup and
the full population to arrive at one of the three conclusions.

We focused on scenarios, where a single
subgroup and possibly the complement were considered for testing
apart from the full population.
Nevertheless the methodology we present here is not restricted
to this setting and can be used to test in an arbitrary number 
of populations. We extend the method to more general scenarios
in the Appendix in Section A.1.
While one could therefore include a large number of subgroups and
use the presented methodology to identify subgroups
with an enhanced effect, we would like to point out that in contrast
to other subgroup identification approaches our method does not test
 for an interaction between subgroup and treatment. A significant
result in a subgroup does therefore not necessarily indicate, 
that the subgroup shows a differential effect and 
could be caused by a treatment effect, that is large but constant
across all populations.
Still, there are certain outcomes, which can be indicative of a subgroup
with a larger treatment effect, most notably, when the population
null hypothesis for the subgroup is rejected but the population
null hypothesis for the full population is not. In our simulations
this can be seen for some scenarios in Table \ref{tab:poa_where_hom_emax}.
For scenarios, where there is only an effect in the subgroup,
the null hypothesis for the subgroup is more often rejected than
for the full population.

We only consider normally distributed outcomes in this article. The MCP-Mod
methodology has been generalized to other types of outcomes
in \cite{pinh:born:glim:2014}. It is possible to apply the presented
methodology to general parametric models. In the general case of
MCP-Mod the approximate multivariate normal distribution is used.
We already use this approximation in this article
in Section \ref{sec:sim_hetvar}  and show that it is a reasonable
approximation for large sample sizes.
The correlation matrices we discuss in Section \ref{sec:fwer}
can simply be used for the multivariate normal instead of the
multivariate $t$ in the general case.

The testing procedure discussed here is a single-step test. Alternatively a 
sequentially rejective test using the closed testing principle could be used as well.
A similar procedure is discussed for the single-populaton MCP-Mod in
\cite{bret:born:koen:2017}. Furthermore, similarly as in \cite{bret:posc:glim:2011},
the multiple testing procedure can be generalized to a weighted test and
extended to a graph-based closed testing procedure. 

The approach we present here only extends the MCP-part of 
MCP-Mod to multiple populations, ignoring the modeling part.
In the standard MCP-Mod either the dose-response model with the lowest
$p$-value is used for dose estimation or all significant models
are combined via model averaging. Similar solutions in the multi-population
scenario could be fitting the model with an interaction term for the subgroup,
if the subgroup is significant, or using a model averaging approach that
also includes subgroups, as for example discussed in \cite{born:ohls:magn:2017}.
Extending the modeling part of the methodology to trials
with multiple populations therefore remains as an area for further research.
\newpage

\section*{Acknowledgements}
\small
 This work was supported by funding from the European Union's Horizon 2020 research and innovation programme under the Marie Sklodowska-Curie grant agreement No 633567 and by the Swiss State Secretariat for Education, Research and Innovation (SERI) under contract number 999754557 . The opinions expressed and arguments employed herein do not necessarily reflect the official views of the Swiss Government.\\

\begin{minipage}{.5\textwidth}
\includegraphics[scale = 0.03]{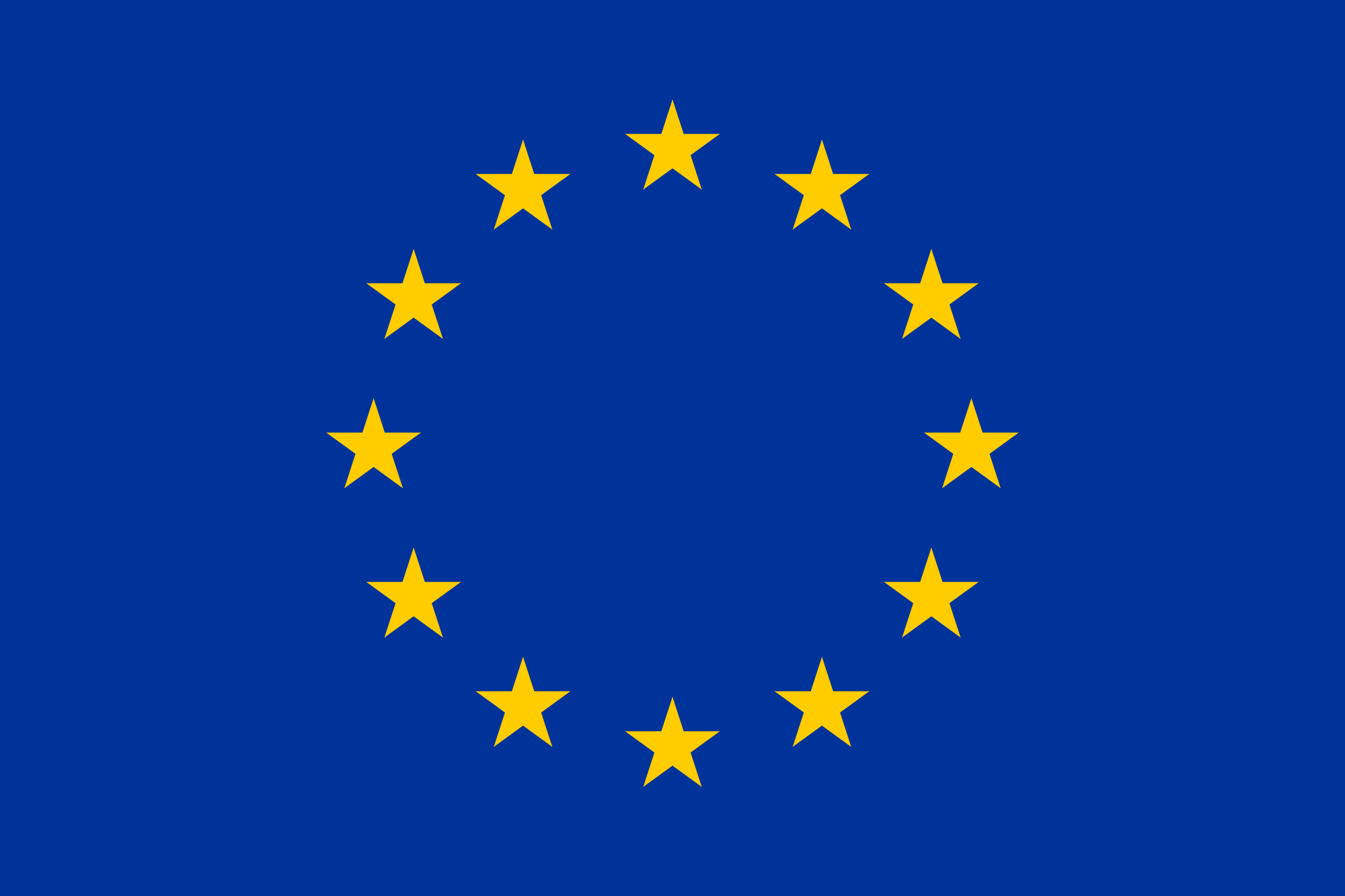}
\end{minipage}%
\begin{minipage}{.5\textwidth}
\includegraphics[scale = 1]{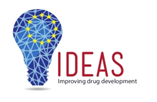}
\end{minipage}

\section*{Appendix}

\subsection*{A.1. Joint distribution in the general case}
\label{sec:app_gen}
In this Section we derive the correlation matrix for the
joint distribution of the contrast tests.
We will derive the matrix for a general case with an arbitrary number of 
populations and also include the possibility that different
candidate shapes are tested in each population. The scenarios described in
\ref{sec:fwer} are special cases of what we discuss in this Section.
 
We assume that we have $W$ populations that
we consider for testing. These populations are contained in the family
$U := \{P_1, \dots, P_W\}$, where each population $P\in U$
is a set of patients, so that $P \subset \{1,\dots,n\}$.
As in the main part of the article we need to introduce a second family of
pairwise disjoint populations $V := \{P_1^{*}, \dots, P_{W^{*}}^{*}\}$,
so that $\bigcup\limits_{r = 1}^{W^{*}} P_r^{*} = \{1, \dots, n\}$,
if we want to allow for populations with different variances.
We then assume that our responses come from the model
\begin{equation}	
\begin{split}
	Y_{ij}^{(P^{*})} =\mu_i^{(P^{*})} + \epsilon_{ij}^{(P^{*})}; 
\epsilon_{ij}^{(P^{*})} \sim N(0, \sigma_{P^{*}}^2) \hspace{0.5cm} i.i.d,\\
 i = 1,...,k, j = 1,...,n_i^{(P^{*})}, P^{*}\in V
\end{split}
\label{eq:drmod2_app}
\end{equation}
and the vector of responses is $\boldsymbol{Y} = (Y_{11},...,Y_{1n_1},...,Y_{K1},...,Y_{Kn_K})$
with $Cov(\boldsymbol{Y}) := \boldsymbol{\Sigma_Y}$.
We therefore allow each population in $V$ to have a different variance. We
need to make the distinction between $V$ and $U$, 
since the populations in $U$ need not necessarily
be pairwise disjoint. For example in the main article, subgroup and complement
are in both families but the full population is only in $U$. 

We now assume that in each population $P\in U$ we test for $Z_P$
different candidate shapes. In each population P we therefore perform tests of
$H_0^{(P, m)}: {c_m^{(P)}}^ {'}
\boldsymbol{\mu^{(P)}} = 0$ against the alternatives $H_1^{(P, m)}: {c_m^{(P)}}^ {'}
\boldsymbol{\mu^{(P)}} > 0$, for $m=1,\dots, Z_P$.
The test statistics for these hypotheses are then
\begin{equation}
T_m^{(P)} = \frac{(\boldsymbol{c_m^{(P)}})^{'}\bar{\boldsymbol{Y}}^{(P)}}{\sqrt{(\hat{\sigma}^{(P)})^2
\sum\limits_{i = 1}^{k} \boldsymbol{c_m^{(P)}}^2/n_i^{(P)}}} \hspace{0.5cm},
\label{eq:teststats_app}
\end{equation}
where $\bar{\boldsymbol{Y}}^{(P)}$ is the vector of the estimated means on each
dose level $\boldsymbol{\bar{Y}^{(P)}} = (\bar{Y_1}^{(P)},...,\bar{Y_k}^{(P)})^{'}$
and $(\hat{\sigma}^{(P)})^2$ the estimated variances.

For the joint distribution of the test statistics we need to derive the correlation
matrix of the vector 
\[
\boldsymbol{T^{*} := [(c_{1}^{(P_1)})^{'}\bar{Y}^{(P_1)},
\dots, (c_{Z_{P_1}}^{(P_1)})^{'}\bar{Y}^{(P_1)}, \dots,
(c_{1}^{(P_W)})^{'}\bar{Y}^{(P_W)},
\dots, (c_{Z_{P_W}}^{(P_W)})^{'}\bar{Y}^{(P_W)}]^{'}}
\] 
We denote by $\Gamma_P := [\boldsymbol{c_1^{(P)}},...,\boldsymbol{c_{Z_P}^{(P)}}]$ the matrix
of optimal contrast coefficients for population $P$ and by $A_P\in\mathbb{R}^{k x n}$
a matrix with entries 
\[
	(a_{ij}) = \begin{cases} \frac{1}{n_i^{(P)}} & \text{if the j-th patient is on dose i and is in population P}\\
	0 & \text{else}\end{cases},
\]
so that $\boldsymbol{A_P}\boldsymbol{Y} = \bar{\boldsymbol{Y}}^{(P)}$.
We can then write the required covariance matrix between all contrasts as 
\[
\boldsymbol{Cov(T^{*}) = 
	Cov\left(\begin{bmatrix}
	\Gamma_{P_1}^{'} & \huge 0 & \huge 0\\
	\huge 0 &  \ddots & \huge 0\\
        \huge 0 &  \huge 0 & \boldsymbol\Gamma_{P_W}^{'}
	\end{bmatrix}
	\begin{bmatrix}
	\bar{Y}^{(P_1)}\\
        \vdots\\
	\bar{Y}^{(P_W)}
	\end{bmatrix}\right) =
	Cov\left(\begin{bmatrix}
	\Gamma_{P_1}^{'} & \huge 0 & \huge 0\\
	\huge 0 &  \ddots & \huge 0\\
        \huge 0 &  \huge 0 & \Gamma_{P_W}^{'}
	\end{bmatrix}
	\begin{bmatrix}
	A_{P_1}\\
	\vdots\\
	A_{P_U}
	\end{bmatrix}
	Y
	\right) =
}
\]
\[
\boldsymbol{
\begin{bmatrix}
	\Gamma_{P_1}^{'}A_{P_1} \\
	\vdots\\
	\Gamma_{P_W}^{'}A_{P_W}
	\end{bmatrix}
	\Sigma_Y
	\begin{bmatrix}
	\Gamma_{P_1}^{'}A_{P_1} \\
	\vdots\\
	\Gamma_{P_W}^{'}A_{P_W}
	\end{bmatrix}^{'} =
	\begin{bmatrix}
	\Gamma_{P_1}^{'}A_{P_1}\Sigma_YA_{P_1}^{'}\Gamma_{P_1}^{'} &
	\hdots &\Gamma_{P_1}^{'}A_{P_1}\Sigma_YA_{P_W}^{'}\Gamma_{P_W}^{'}\\
	\vdots & \ddots & \vdots\\
	\Gamma_{P_W}^{'}A_{P_W}\Sigma_YA_{P_1}^{'} \Gamma_{P_1}^{'} &
	\hdots & \Gamma_{P_W}^{'}A_{P_W}\Sigma_YA_{P_W}^{'}\Gamma_{P_W}^{'}\\
	\end{bmatrix} 
} :=
\]
\[
\boldsymbol{
	\begin{bmatrix}
	\Sigma_{P_1P_1} & \hdots & \Sigma_{P_1P_W} \\
	\vdots & \ddots & \vdots\\
	\Sigma_{P_WP_1} & \hdots & \Sigma_{P_WP_W} \\
	\end{bmatrix}.}
\]
From this covariance matrix we can obtain the
required correlation matrix $\boldsymbol{R}$.
\[
\boldsymbol{
 	R=
	\begin{bmatrix}
	R_{P_1P_1} & \hdots & R_{P_1P_W} \\
	\vdots & \ddots & \vdots\\
	R_{P_WP_1} & \hdots & R_{P_WP_W} \\
	\end{bmatrix}},
\]
where the entries of the submatrix $\boldsymbol{R_{PQ}}$ for the
correlations between tests in populations $P$ and $Q$ 
are
\begin{gather*}
	(\rho_{ij})^{(PQ)} = 
	\frac{\sum\limits_{l =1}^K
         [c_{il}c_{jl}/(n_l^{(P)}n_l^{(Q)})
	\sum\limits_{P^{*}\in V} 
	n_l^{(P \cap Q\cap P^{*})}\sigma_{P^{*}}^2]}
	{\sqrt{\sum\limits_{l =1}^K[{c_{il}}^2/{n_l^{(P)}}^2\sum\limits_{P^{*}\in V} 
	n_l^{(P \cap P^{*})}\sigma_{P^{*}}^2] 
	\sum\limits_{l =1}^K[{c_{jl}}^2/{n_l^{(Q)}}^2\sum\limits_{P^{*}\in V} 
	n_l^{(Q \cap P^{*})}\sigma_{P^{*}}^2}]}.
\end{gather*}
The cases discussed in the main part of this article correspond
to $U =\{F, S, C\}$ and $V = \{S, C\}$. 

As in the main part of the article the form of the joint distribution, which
is used to derive critical values depends on the variance estimators in
(\ref{eq:teststats_app}). Possible joint distributions are summarized in
Table \ref{tab:methods}. In the general case the degrees of freedom have
to be adjusted for the possible higher number of populations as well.

\subsection*{A.2. Simulation results for model selection}
Table \ref{tab:modsec_hom} shows the simulation results for
model selection in the case of homogeneous variances in subgroup and
complement. The Table shows the probability of choosing the correct model
as the best model in the MCP-step. The best model is defined as
the model with the lowest $p$-value over all tested populations.

\begin{table}[ht]
\centering
\caption{Probability of choosing the correct dose-response model for single population (SP) and multi-population (MP) testing methods. The best model is chosen based on the lowest significant $p$-value across all populations. 
Probabilities shown are conditional on the global null hypothesis being rejected.
}
\small
\begin{tabular}{ll|lllllllll}
\hline
& & \multicolumn{8}{c}{Scenario and Prevalence}\\
& & \textit{same} & \textit{same} & \textit{same} & \textit{double} & \textit{double} & \textit{double} & \textit{only} & \textit{only} & \textit{only} \\  
Correct model & Method & 0.25 & 0.5 & 0.75 & 0.25 & 0.5 & 0.75 & 0.25 & 0.5 & 0.75 \\ 
  \hline
emax & SP & 0.50 & 0.51 & 0.50 & 0.37 & 0.44 & 0.45 & 0.26 & 0.33 & 0.40 \\ 
  & MP-Pooled(F+S) & 0.49 & 0.49 & 0.48 & 0.38 & 0.44 & 0.45 & 0.34 & 0.40 & 0.45 \\ 
  & MP-Pooled(F+S+C) & 0.48 & 0.48 & 0.49 & 0.37 & 0.43 & 0.46 & 0.34 & 0.41 & 0.46 \\ 
  linear & SP & 0.27 & 0.28 & 0.27 & 0.17 & 0.21 & 0.24 & 0.10 & 0.16 & 0.21 \\ 
  & MP-Pooled(F+S) & 0.27 & 0.27 & 0.27 & 0.17 & 0.21 & 0.24 & 0.18 & 0.20 & 0.24 \\ 
  & MP-Pooled(F+S+C) & 0.27 & 0.28 & 0.27 & 0.17 & 0.20 & 0.24 & 0.19 & 0.20 & 0.24 \\ 
  exponential & SP & 0.69 & 0.69 & 0.69 & 0.60 & 0.62 & 0.66 & 0.44 & 0.54 & 0.63 \\ 
  & MP-Pooled(F+S) & 0.68 & 0.69 & 0.68 & 0.58 & 0.62 & 0.67 & 0.54 & 0.60 & 0.65 \\ 
  & MP-Pooled(F+S+C) & 0.68 & 0.67 & 0.67 & 0.57 & 0.61 & 0.67 & 0.52 & 0.61 & 0.65 \\ 
  logistic & SP & 0.56 & 0.56 & 0.57 & 0.43 & 0.48 & 0.52 & 0.30 & 0.40 & 0.48 \\ 
  & MP-Pooled(F+S) & 0.56 & 0.55 & 0.57 & 0.42 & 0.48 & 0.52 & 0.38 & 0.46 & 0.52 \\ 
  & MP-Pooled(F+S+C) & 0.55 & 0.55 & 0.57 & 0.41 & 0.48 & 0.52 & 0.38 & 0.46 & 0.53 \\ 
  quadratic & SP & 0.76 & 0.76 & 0.76 & 0.64 & 0.69 & 0.71 & 0.43 & 0.60 & 0.67 \\ 
  & MP-Pooled(F+S) & 0.75 & 0.75 & 0.75 & 0.62 & 0.68 & 0.71 & 0.52 & 0.67 & 0.71 \\ 
  & MP-Pooled(F+S+C) & 0.74 & 0.75 & 0.74 & 0.62 & 0.68 & 0.71 & 0.51 & 0.66 & 0.71 \\  
\hline
\end{tabular}
\label{tab:modsec_hom}
\end{table}

\FloatBarrier
\bibliographystyle{wileyj}

\bibliography{bibl}

\newpage 
\begin{center}
\textbf{\huge Supporting Information}
\end{center}

\setcounter{section}{0}

\section{Additional simulation results for homoscedastic scenarios}

\subsection{Results for other data-generating models}

\begin{table}[ht]
\scriptsize
\centering
\begin{tabular}{l|rrrr}
Method & global & F & S & C \\
  \hline
SP & 0.05 & 0.05 & - & - \\ 
MP-Pooled(F+S) & 0.05 & 0.03 & 0.03 & -\\  
MP-Pooled(F+S+C)  & 0.05 & 0.03 & 0.02 & 0.02 \\ 
MP-MultDF(F+S)  & 0.05 & 0.03 & 0.02 & -  \\ 
MP-MultDF(F+S+C) & 0.05 & 0.02 & 0.02 & 0.02 \\ 

\end{tabular}
\caption{Probability to reject the global null hypothesis and 
population null hypotheses for single population (SP) and
multi-population (MP) testing methods. Data
are generated from a constant model under homoscedasticity.}
\label{tab:poa_where_hom_const}
\end{table}

\begin{table}[ht]
\scriptsize
\centering
\begin{tabular}{ll|rrrrrrrrr}
& & \multicolumn{8}{c}{Scenario and Prevalence}\\
& & \textit{same} & \textit{same} & \textit{same} & \textit{double} & \textit{double} & \textit{double} & \textit{only} & \textit{only} & \textit{only} \\  
Method & Population & 0.25 & 0.5 & 0.75 & 0.25 & 0.5 & 0.75 & 0.25 & 0.5 & 0.75 \\ 
  \hline
SP & global & 0.88 & 0.88 & 0.87 & 0.54 & 0.67 & 0.79  & 0.16 & 0.40 & 0.66 \\ 
  MP-Pooled(F+S) & global & 0.82 & 0.84 & 0.85  & 0.54 & 0.69 & 0.80  & 0.33 & 0.59 & 0.75 \\ 
   & F & 0.81 & 0.82 & 0.83  & 0.43 & 0.58 & 0.73 & 0.10 & 0.32 & 0.60 \\ 
   & S & 0.30 & 0.55 & 0.72  & 0.31 & 0.54 & 0.73 & 0.29 & 0.54 & 0.72 \\ 
  MP-Pooled(F+S+C) & global & 0.82 & 0.82 & 0.82  & 0.51 & 0.64 & 0.75 & 0.30 & 0.52 & 0.69 \\ 
   & F & 0.78 & 0.78 & 0.78  & 0.39 & 0.52 & 0.67 & 0.09 & 0.26 & 0.52\\ 
   & S & 0.27 & 0.49 & 0.65  & 0.28 & 0.48 & 0.66 & 0.26 & 0.48 & 0.65 \\ 
   & C & 0.65 & 0.46 & 0.26  & 0.20 & 0.14 & 0.09 & 0.02 & 0.02 & 0.02 \\ 
  MP-MultDF(F+S) & global & 0.83 & 0.85 & 0.87  & 0.53 & 0.69 & 0.81 & 0.35 & 0.58 & 0.75 \\ 
   & F & 0.81 & 0.83 & 0.85  & 0.43 & 0.59 & 0.74 & 0.11 & 0.31 & 0.59 \\ 
   & S & 0.28 & 0.55 & 0.74  & 0.29 & 0.55 & 0.72 & 0.31 & 0.54 & 0.72 \\ 
  MP-MultDF(F+S+C) & global & 0.83 & 0.83 & 0.83 & 0.50 & 0.65 & 0.75  & 0.31 & 0.52 & 0.68\\ 
   & F & 0.78 & 0.79 & 0.79  & 0.39 & 0.53 & 0.67 & 0.09 & 0.26 & 0.51 \\ 
   & S & 0.25 & 0.49 & 0.66 & 0.26 & 0.49 & 0.65  & 0.27 & 0.48 & 0.65 \\ 
   & C & 0.66 & 0.48 & 0.26  & 0.20 & 0.13 & 0.09 & 0.02 & 0.02 & 0.02 \\

\end{tabular}
\caption{Probability to reject the global null hypothesis and 
population null hypotheses for single population (SP) and
multi-population (MP) testing methods. Data
are generated from a linear model under homoscedasticity.}
\label{tab:poa_where_hom_lin}
\end{table}

\begin{table}[ht]
\scriptsize
\centering
\begin{tabular}{ll|rrrrrrrrr}
& & \multicolumn{8}{c}{Scenario and Prevalence}\\
& & \textit{same} & \textit{same} & \textit{same} & \textit{double} & \textit{double} & \textit{double} & \textit{only} & \textit{only} & \textit{only} \\ 
Method & Population & 0.25 & 0.5 & 0.75 & 0.25 & 0.5 & 0.75 & 0.25 & 0.5 & 0.75 \\ 
  \hline
SP & global & 0.86 & 0.86 & 0.86  & 0.51 & 0.65 & 0.76 & 0.16 & 0.36 & 0.64 \\ 
  MP-Pooled(F+S) & global & 0.80 & 0.82 & 0.84  & 0.50 & 0.66 & 0.78 & 0.31 & 0.55 & 0.73 \\ 
   & F & 0.78 & 0.80 & 0.82  & 0.41 & 0.56 & 0.70 & 0.10 & 0.29 & 0.58\\ 
   & S & 0.26 & 0.52 & 0.70 & 0.28 & 0.51 & 0.70  & 0.27 & 0.51 & 0.69 \\ 
  MP-Pooled(F+S+C) & global & 0.80 & 0.80 & 0.80 & 0.47 & 0.61 & 0.72  & 0.28 & 0.49 & 0.67 \\ 
   & F & 0.75 & 0.75 & 0.76  & 0.37 & 0.50 & 0.63 & 0.08 & 0.24 & 0.49\\ 
   & S & 0.23 & 0.45 & 0.63 & 0.24 & 0.44 & 0.63  & 0.24 & 0.45 & 0.63 \\ 
   & C & 0.62 & 0.43 & 0.23  & 0.18 & 0.14 & 0.09 & 0.02 & 0.02 & 0.02 \\ 
  MP-MultDF(F+S) & global & 0.81 & 0.82 & 0.85 & 0.49 & 0.66 & 0.79  & 0.30 & 0.55 & 0.74 \\ 
   & F & 0.80 & 0.80 & 0.83 & 0.40 & 0.56 & 0.71  & 0.10 & 0.28 & 0.58 \\ 
   & S & 0.27 & 0.52 & 0.71  & 0.26 & 0.50 & 0.71 & 0.26 & 0.51 & 0.71 \\ 
  MP-MultDF(F+S+C) & global & 0.81 & 0.81 & 0.81  & 0.46 & 0.61 & 0.72 & 0.26 & 0.49 & 0.67 \\ 
   & F & 0.76 & 0.76 & 0.77  & 0.36 & 0.50 & 0.64 & 0.08 & 0.23 & 0.49 \\ 
   & S & 0.24 & 0.46 & 0.64  & 0.23 & 0.44 & 0.63 & 0.22 & 0.45 & 0.63 \\ 
   & C & 0.62 & 0.44 & 0.24  & 0.18 & 0.13 & 0.09 & 0.02 & 0.02 & 0.02 \\ 

\end{tabular}
\caption{Probability to reject the global null hypothesis and 
population null hypotheses for single population (SP) and
multi-population (MP) testing methods. Data
are generated from an exponential model under homoscedasticity.}
\label{tab:poa_where_hom_expo}
\end{table}

\begin{table}[ht]
\scriptsize
\centering
\begin{tabular}{ll|rrrrrrrrr}
& & \multicolumn{8}{c}{Scenario and Prevalence}\\
& & \textit{same} & \textit{same} & \textit{same} & \textit{double} & \textit{double} & \textit{double} & \textit{only} & \textit{only} & \textit{only} \\ 
Method & Population & 0.25 & 0.5 & 0.75 & 0.25 & 0.5 & 0.75 & 0.25 & 0.5 & 0.75 \\ 
  \hline
SP & global & 0.95 & 0.95 & 0.95 & 0.64 & 0.80 & 0.88  & 0.20 & 0.49 & 0.78\\ 
  MP-Pooled(F+S) & global & 0.92 & 0.93 & 0.94  & 0.64 & 0.81 & 0.90 & 0.42 & 0.70 & 0.87 \\ 
   & F & 0.92 & 0.92 & 0.93 & 0.54 & 0.72 & 0.84  & 0.13 & 0.41 & 0.73\\ 
   & S & 0.38 & 0.66 & 0.85  & 0.39 & 0.67 & 0.85 & 0.38 & 0.67 & 0.85\\ 
  MP-Pooled(F+S+C) & global & 0.92 & 0.92 & 0.92  & 0.61 & 0.77 & 0.86& 0.38 & 0.65 & 0.82 \\ 
   & F & 0.90 & 0.90 & 0.90  & 0.50 & 0.67 & 0.80& 0.11 & 0.36 & 0.66 \\ 
   & S & 0.34 & 0.60 & 0.80  & 0.35 & 0.61 & 0.80& 0.34 & 0.61 & 0.79 \\ 
   & C & 0.80 & 0.60 & 0.35 & 0.27 & 0.19 & 0.11 & 0.02 & 0.02 & 0.02 \\ 
  MP-MultDF(F+S) & global & 0.92 & 0.93 & 0.94  & 0.63 & 0.81 & 0.90& 0.41 & 0.72 & 0.86 \\ 
   & F & 0.92 & 0.92 & 0.93 & 0.53 & 0.72 & 0.85 & 0.13 & 0.41 & 0.73 \\ 
   & S & 0.39 & 0.68 & 0.84  & 0.37 & 0.67 & 0.84& 0.37 & 0.68 & 0.84 \\ 
  MP-MultDF(F+S+C) & global & 0.92 & 0.92 & 0.92 & 0.60 & 0.76 & 0.86 & 0.37 & 0.66 & 0.81 \\ 
   & F & 0.90 & 0.90 & 0.90& 0.49 & 0.66 & 0.80 & 0.11 & 0.35 & 0.66  \\ 
   & S & 0.34 & 0.62 & 0.78  & 0.32 & 0.61 & 0.79& 0.33 & 0.62 & 0.79 \\ 
   & C & 0.80 & 0.60 & 0.34 & 0.25 & 0.20 & 0.11 & 0.02 & 0.02 & 0.02 \\  

\end{tabular}
\caption{Probability to reject the global null hypothesis and 
population null hypotheses for single population (SP) and
multi-population (MP) testing methods. Data
are generated from a logistic model under homoscedasticity.}
\label{tab:poa_where_hom_log}
\end{table}

\begin{table}[ht]
\scriptsize
\centering
\begin{tabular}{ll|rrrrrrrrr}
& & \multicolumn{8}{c}{Scenario and Prevalence}\\
& & \textit{same} & \textit{same} & \textit{same} & \textit{double} & \textit{double} & \textit{double} & \textit{only} & \textit{only} & \textit{only}\\  
Method & Population & 0.25 & 0.5 & 0.75 & 0.25 & 0.5 & 0.75 & 0.25 & 0.5 & 0.75 \\ 
  \hline
SP & global & 0.79 & 0.80 & 0.79 & 0.43 & 0.57 & 0.68 & 0.14 & 0.31 & 0.54 \\ 
  MP-Pooled(F+S) & global & 0.73 & 0.75 & 0.77  & 0.42 & 0.58 & 0.70 & 0.26 & 0.47 & 0.65\\ 
   & F & 0.71 & 0.72 & 0.74  & 0.33 & 0.48 & 0.62& 0.08 & 0.24 & 0.48 \\ 
   & S & 0.22 & 0.42 & 0.61 & 0.23 & 0.43 & 0.62  & 0.23 & 0.43 & 0.61\\ 
  MP-Pooled(F+S+C) & global & 0.72 & 0.71 & 0.72  & 0.39 & 0.53 & 0.64& 0.24 & 0.41 & 0.57 \\ 
   & F & 0.67 & 0.66 & 0.67  & 0.29 & 0.42 & 0.54& 0.07 & 0.20 & 0.40 \\ 
   & S & 0.19 & 0.37 & 0.53 & 0.20 & 0.37 & 0.54 & 0.20 & 0.37 & 0.54 \\ 
   & C & 0.53 & 0.36 & 0.21  & 0.15 & 0.11 & 0.07& 0.02 & 0.02 & 0.02 \\ 
  MP-MultDF(F+S) & global & 0.72 & 0.74 & 0.78  & 0.42 & 0.58 & 0.69 & 0.25 & 0.47 & 0.65\\ 
   & F & 0.70 & 0.72 & 0.74  & 0.33 & 0.48 & 0.61& 0.08 & 0.24 & 0.49 \\ 
   & S & 0.23 & 0.42 & 0.62  & 0.23 & 0.44 & 0.60& 0.22 & 0.43 & 0.61 \\ 
  MP-MultDF(F+S+C) & global & 0.72 & 0.72 & 0.73  & 0.40 & 0.53 & 0.62& 0.23 & 0.41 & 0.58 \\ 
   & F & 0.66 & 0.67 & 0.67  & 0.29 & 0.42 & 0.53 & 0.07 & 0.20 & 0.40\\ 
   & S & 0.20 & 0.36 & 0.54  & 0.19 & 0.38 & 0.52& 0.19 & 0.37 & 0.54 \\ 
   & C & 0.52 & 0.36 & 0.20 & 0.14 & 0.11 & 0.07  & 0.02 & 0.02 & 0.03\\
\end{tabular}
\caption{Probability to reject the global null hypothesis and 
population null hypotheses for single population (SP) and
multi-population (MP) testing methods. Data
are generated from a quadratic model under homoscedasticity.}
\label{tab:poa_where_hom_quad}
\end{table}

\clearpage

\subsection{Detailed power curves}

\begin{figure}[h!]
\centering
\includegraphics[width=\textwidth]{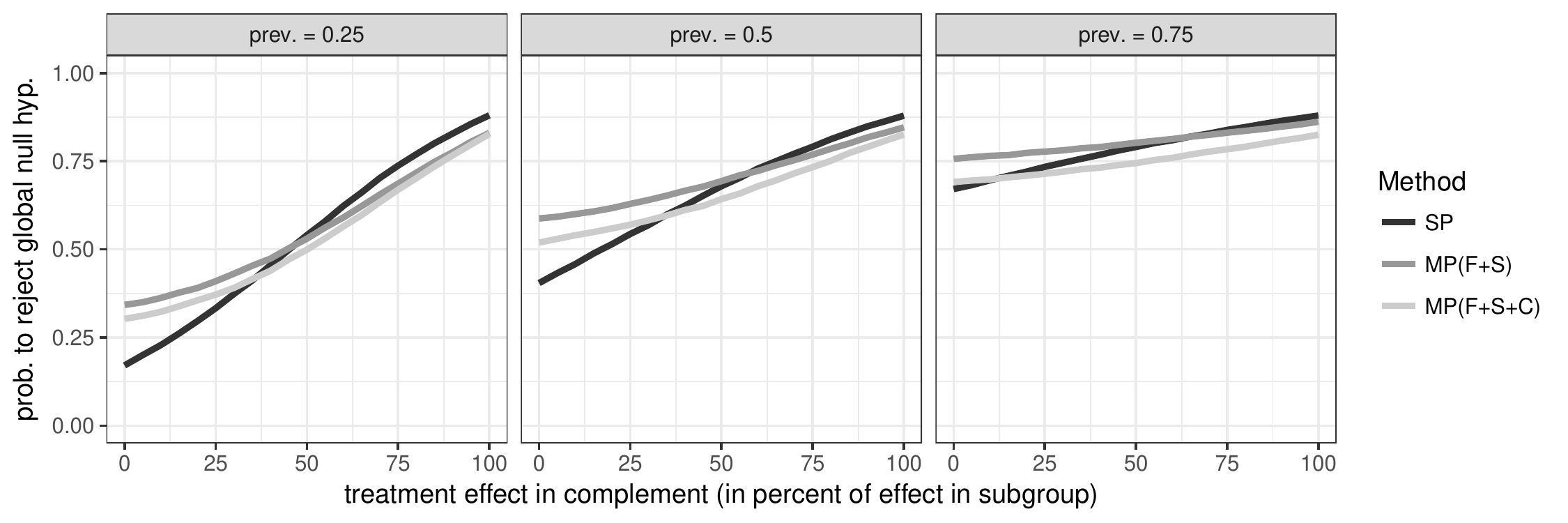}
\caption{Probability to reject the global null hypothesis for single population (SP) and
multi-population (MP) testing methods plotted against the treatment effect in the complement
in relation to the subgroup. Data
are generated from an Emax model under homoscedasticity.}
\label{fig:poa_effpower_emax}
\end{figure}
\begin{figure}[h!]
\centering
\includegraphics[width=\textwidth]{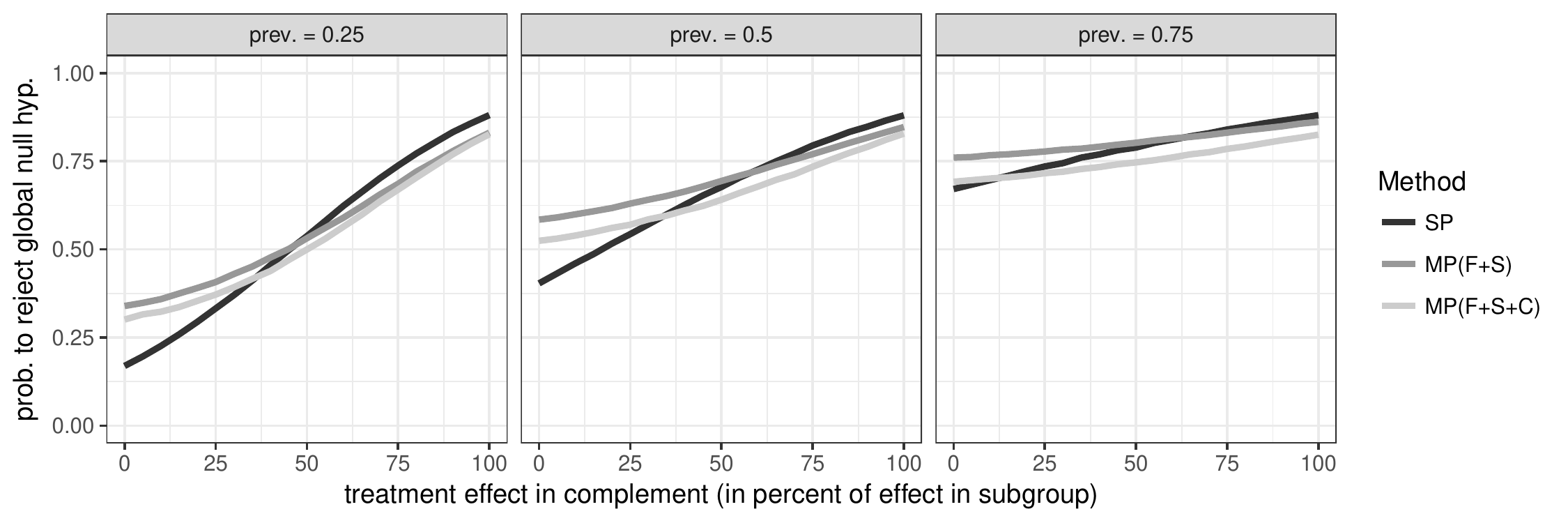}
\caption{Probability to reject the global null hypothesis for single population (SP) and
multi-population (MP) testing methods plotted against the treatment effect in the complement
in relation to the subgroup. Data
are generated from a linear model under homoscedasticity.}
\label{fig:poa_effpower_lin}
\end{figure}
\begin{figure}[h!]
\centering
\includegraphics[width=\textwidth]{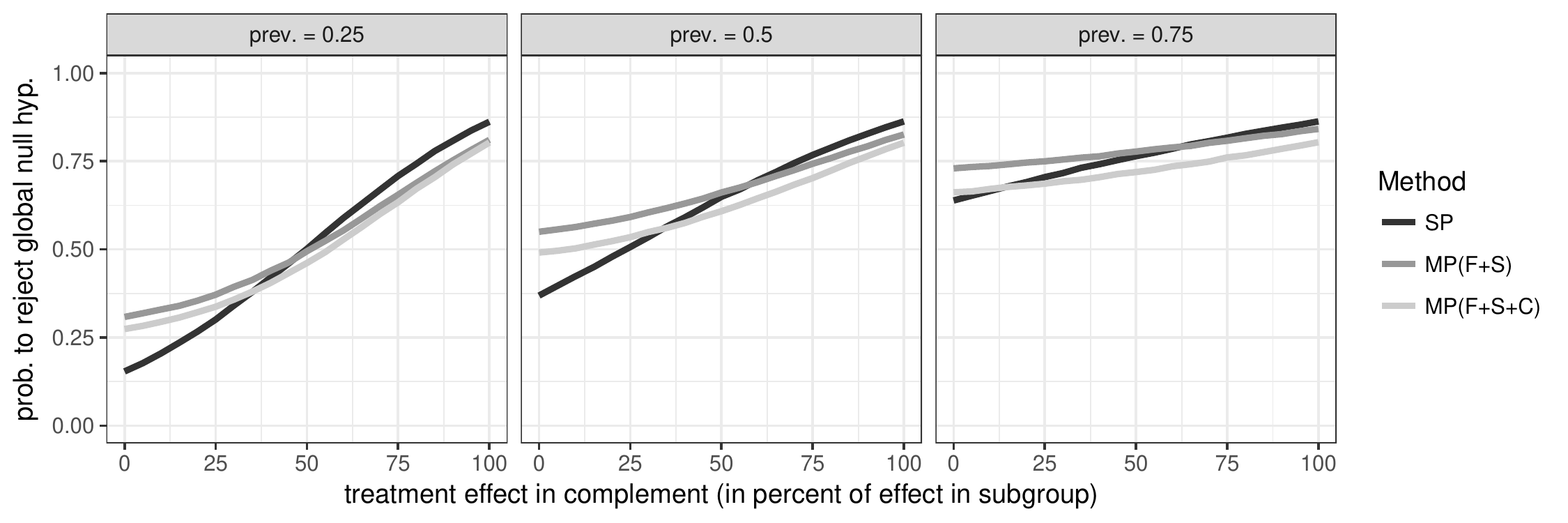}
\caption{Probability to reject the global null hypothesis for single population (SP) and
multi-population (MP) testing methods plotted against the treatment effect in the complement
in relation to the subgroup. Data
are generated from an exponential model under homoscedasticity.}
\label{fig:poa_effpower_expo}
\end{figure}
\begin{figure}[h!]
\centering
\includegraphics[width=\textwidth]{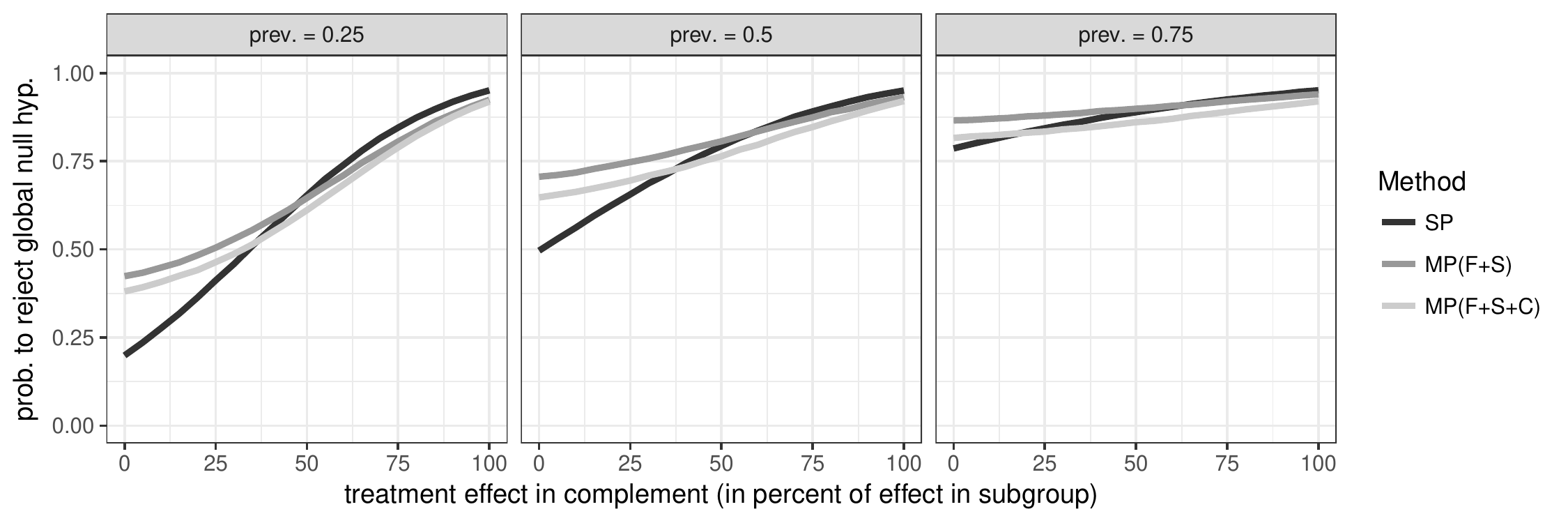}
\caption{Probability to reject the global null hypothesis for single population (SP) and
multi-population (MP) testing methods plotted against the treatment effect in the complement
in relation to the subgroup.  Data
are generated from a logistic model under homoscedasticity.}
\label{fig:poa_effpower_log}
\end{figure}
\begin{figure}[h!]
\centering
\includegraphics[width=\textwidth]{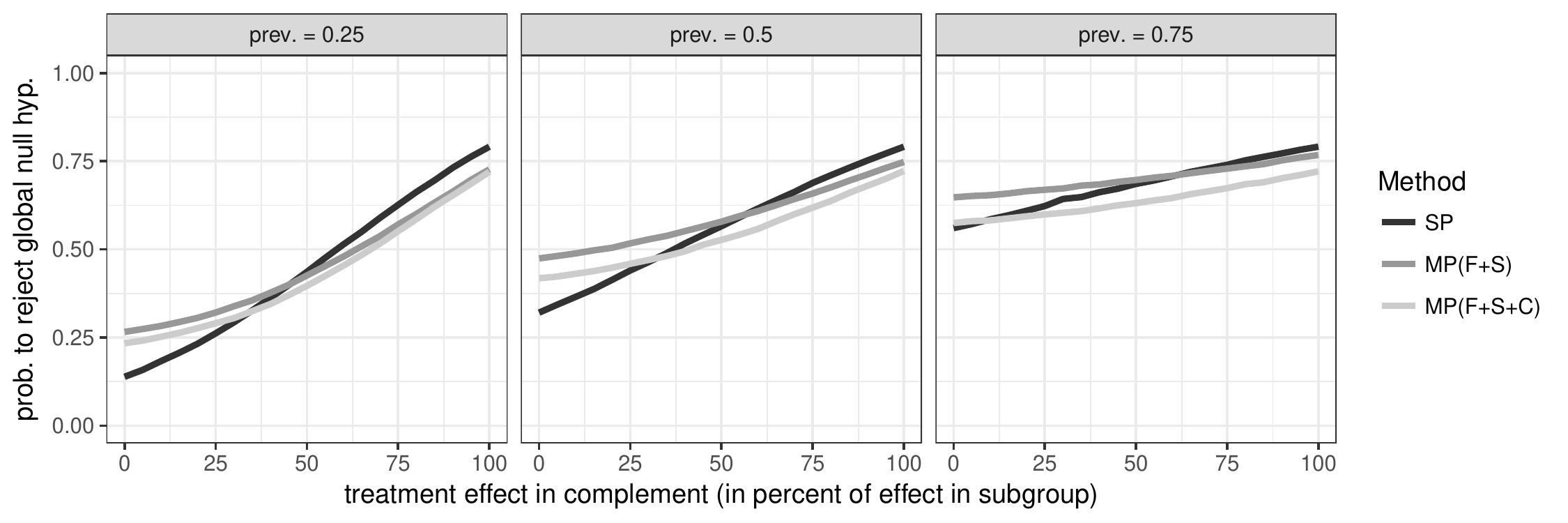}
\caption{Probability to reject the global null hypothesis for single population (SP) and
multi-population (MP) testing methods plotted against the treatment effect in the complement
in relation to the subgroup.  Data
are generated from a quadratic model under homoscedasticity.}
\label{fig:poa_effpower_quad}
\end{figure}
\clearpage
\section{Additional simulation results for heteroscedastic scenarios}

\subsection{Comparison of approximate tests for other data-generating models}
\begin{figure}[h!]
\centering
\includegraphics[width=\textwidth]{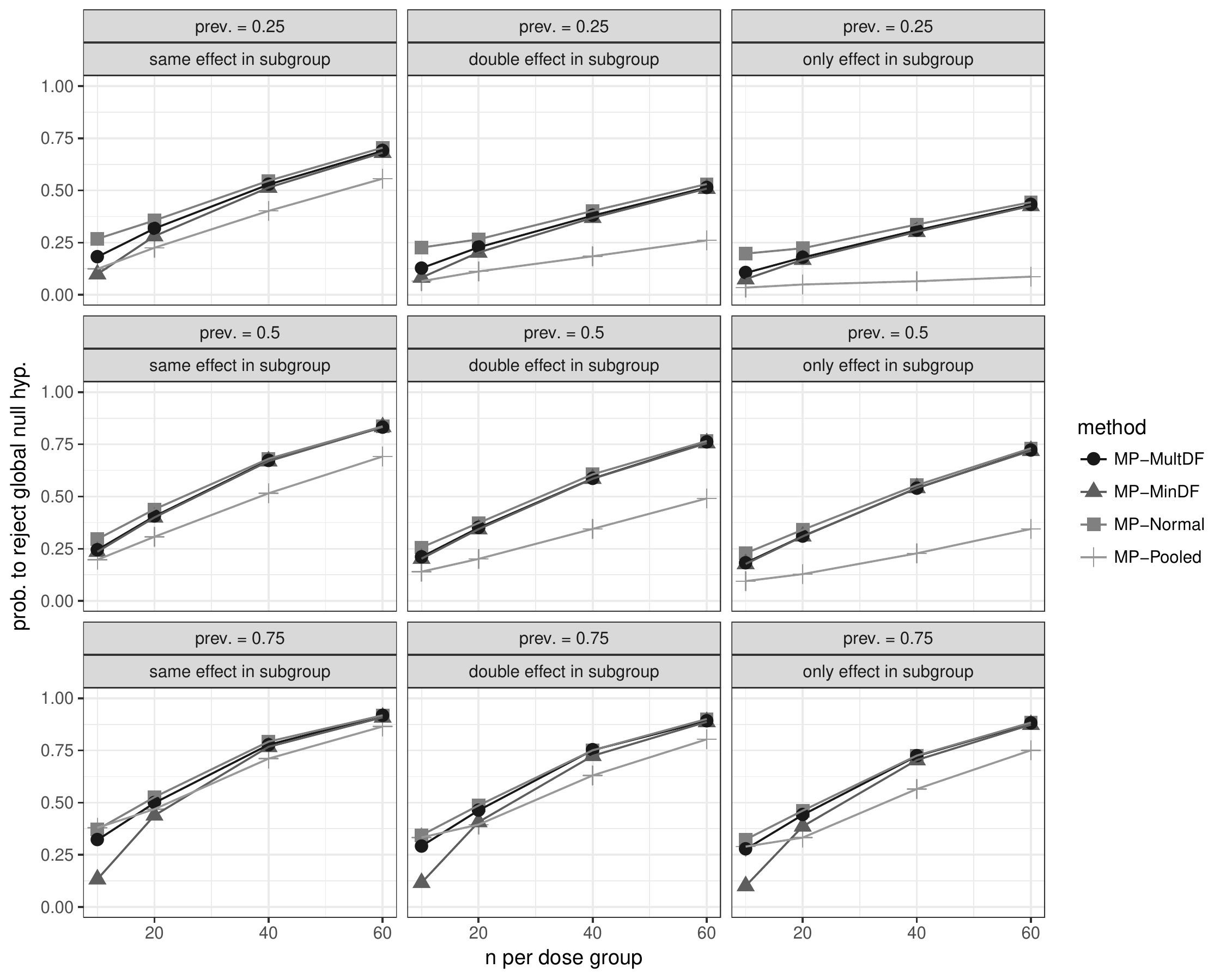}
\caption{Probability to reject the global null hypothesis for single population (SP) and
multi-population (MP) testing methods. Data
are generated from a linear model under heteroscedasticity.
MP-MultDF is used to approximate the joint distribution for
MP testing methods.}
\label{fig:poa_het_lin}
\end{figure}
\begin{figure}[h!]
\centering
\includegraphics[width=\textwidth]{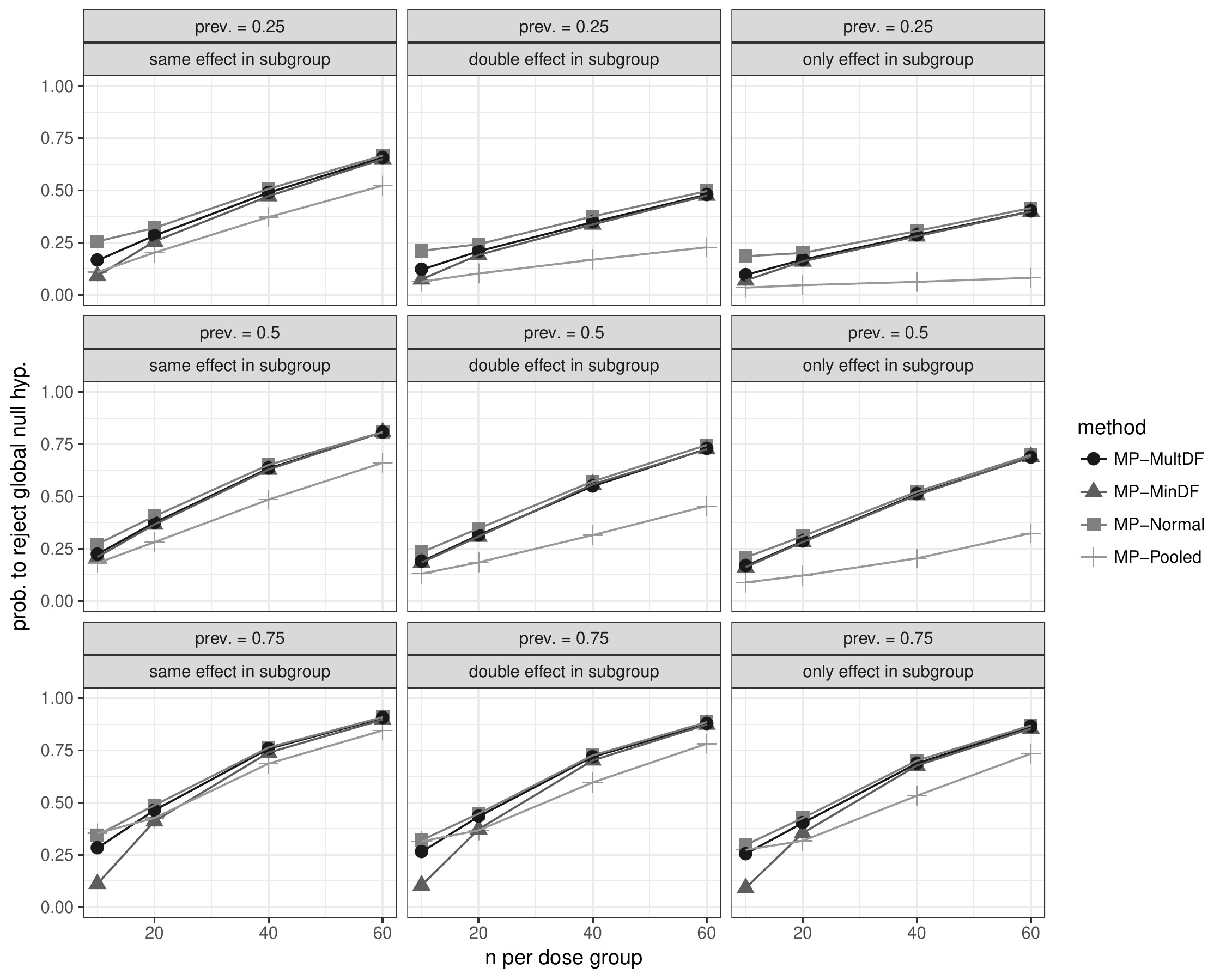}
\caption{Probability to reject the global null hypothesis for single population (SP) and
multi-population (MP) testing methods. Data
are generated from an exponential model under heteroscedasticity.
MP-MultDF is used to approximate the joint distribution for
MP testing methods.}
\label{fig:poa_het_expo}
\end{figure}
\begin{figure}[h!]
\centering
\includegraphics[width=\textwidth]{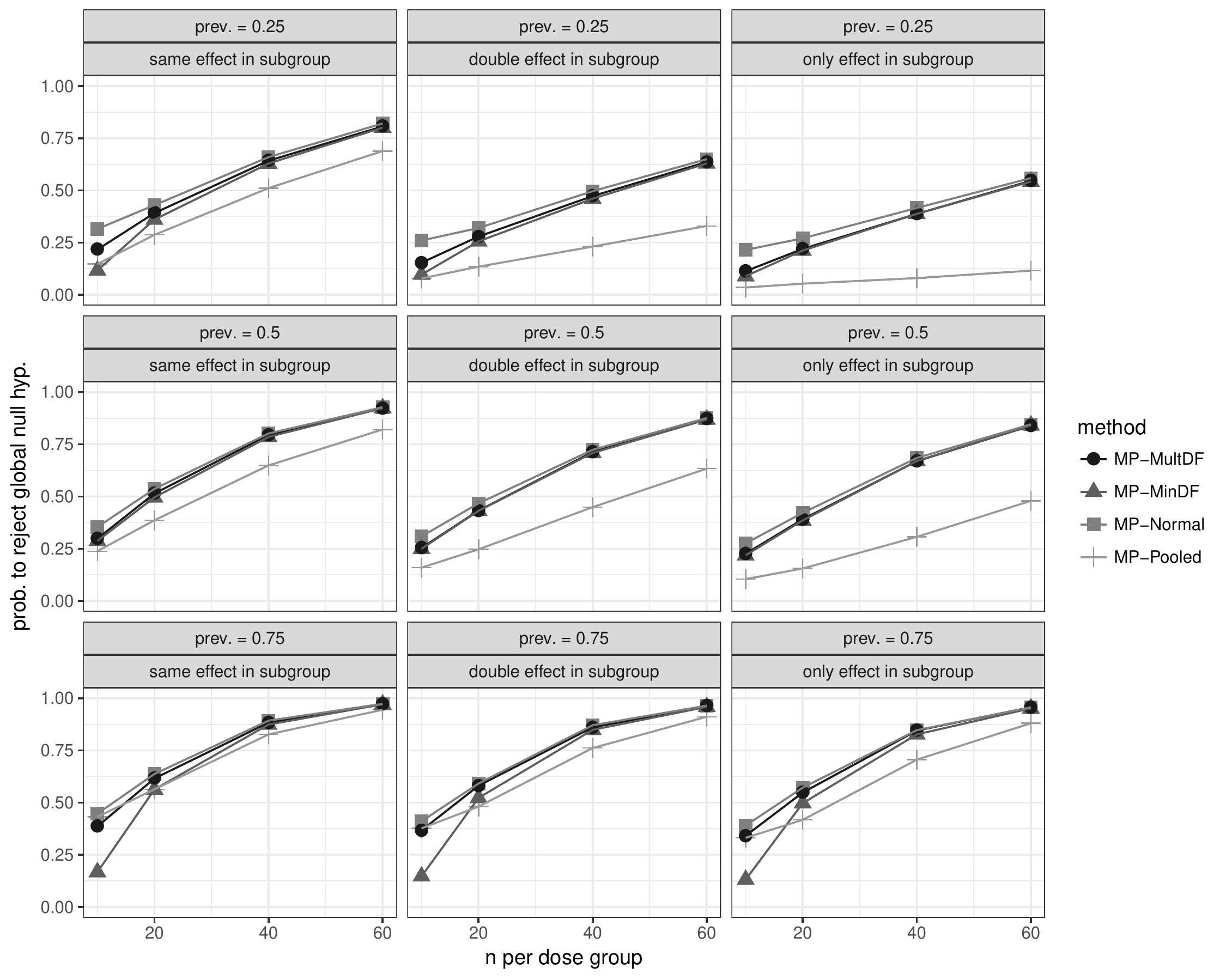}
\caption{Probability to reject the global null hypothesis for single population (SP) and
multi-population (MP) testing methods. Data
are generated from a logistic model under heteroscedasticity.
MP-MultDF is used to approximate the joint distribution for
MP testing methods.}
\label{fig:poa_het_log}
\end{figure}
\begin{figure}[h!]
\centering
\includegraphics[width=\textwidth]{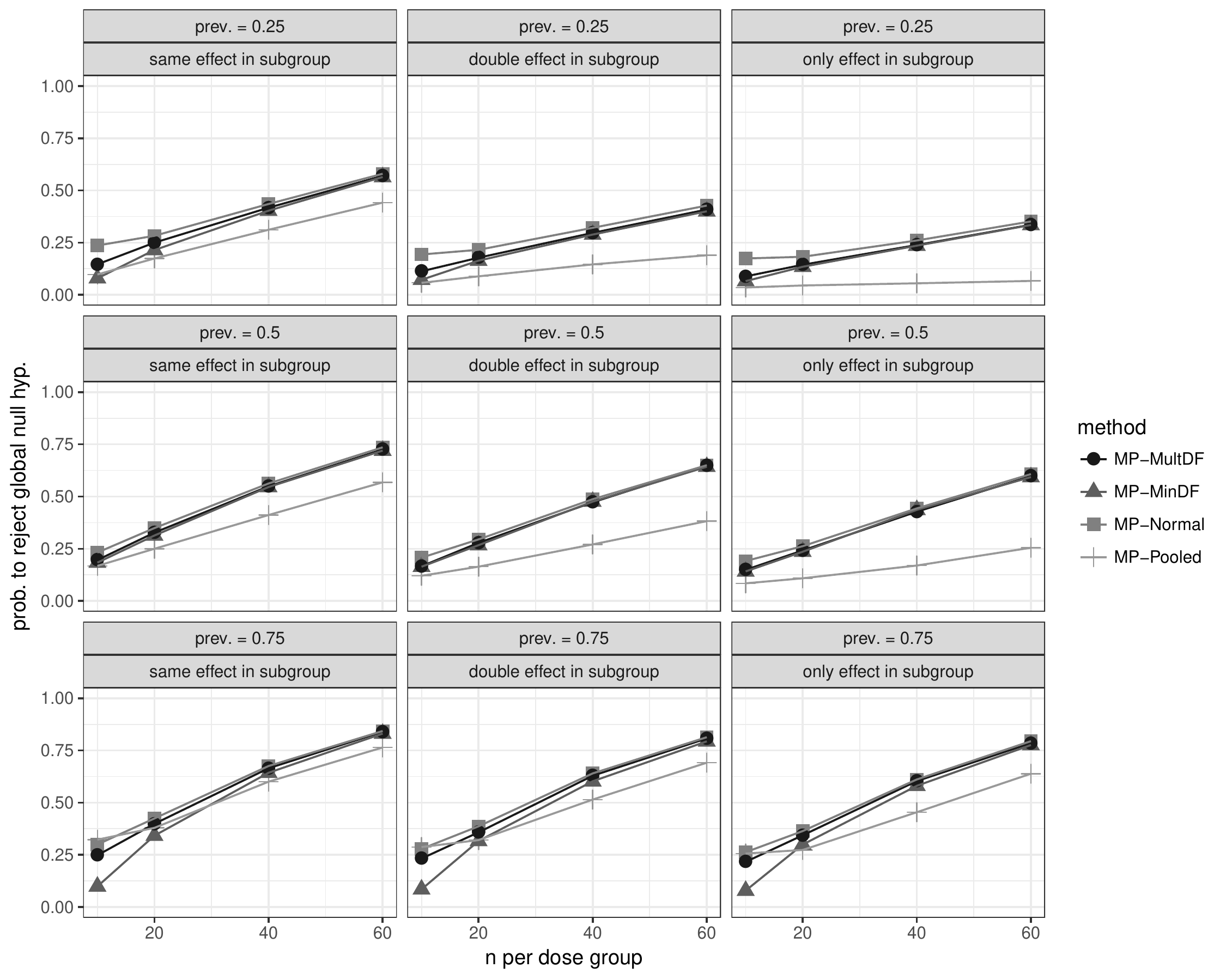}
\caption{Probability to reject the global null hypothesis for single population (SP) and
multi-population (MP) testing methods. Data
are generated from a quadratic model under heteroscedasticity.
MP-MultDF is used to approximate the joint distribution for
MP testing methods.}
\label{fig:poa_het_quad}
\end{figure}

\clearpage

\subsection{Comparison of multi-population MCP to single population MCP for other data-generating models}

\begin{figure}[h!]
\centering
\includegraphics[width=\textwidth]{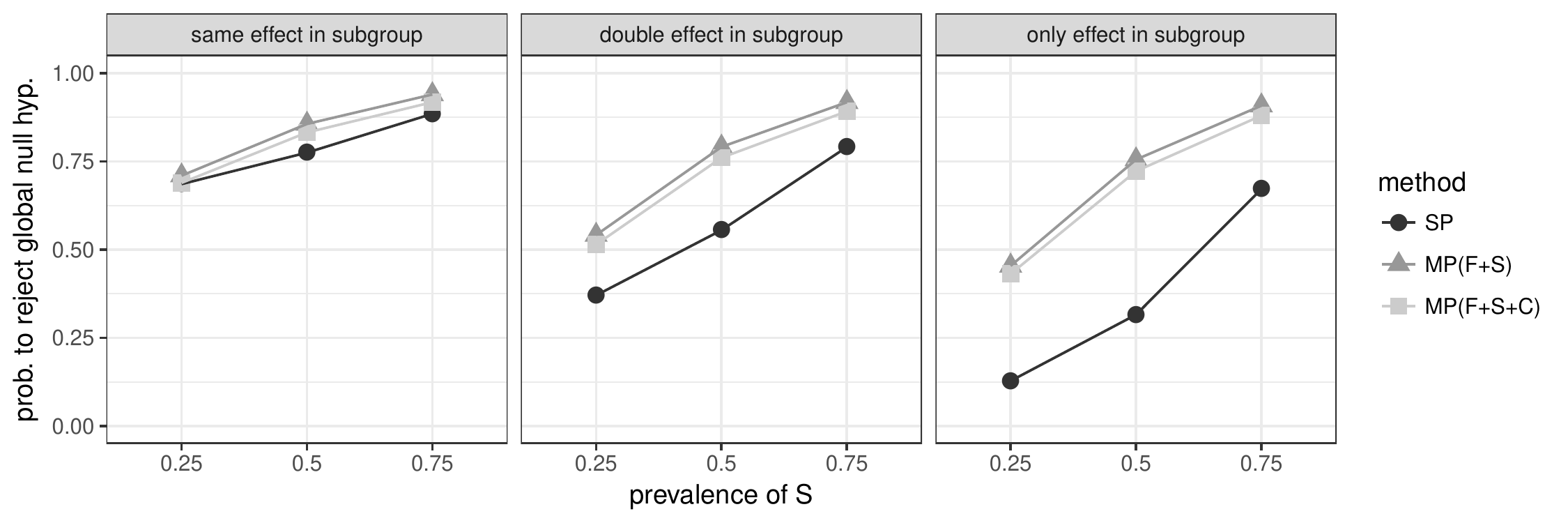}
\caption{Probability to reject the global null hypothesis for single population (SP) and
multi-population (MP) testing methods. Data
are generated from a linear model under heteroscedasticity.
MP-MultDF is used to approximate the joint distribution for
MP testing methods.}
\label{fig:poa_het_lin2}
\end{figure}
\begin{figure}[h!]
\centering
\includegraphics[width=\textwidth]{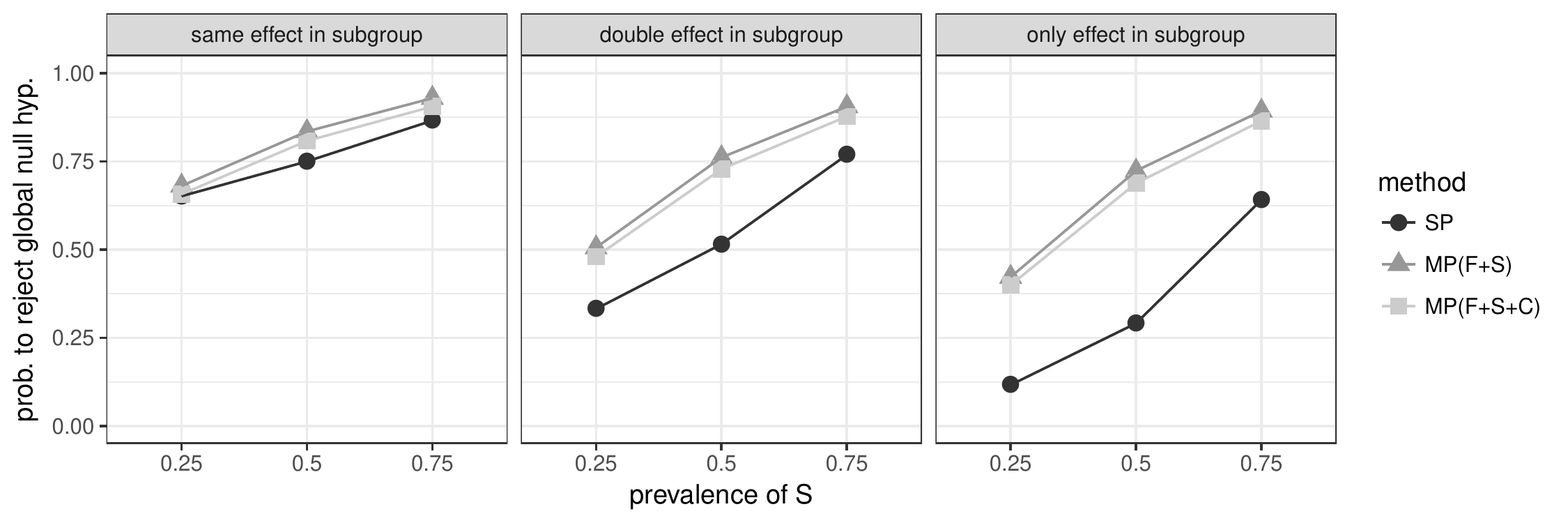}
\caption{Probability to reject the global null hypothesis for single population (SP) and
multi-population (MP) testing methods. Data
are generated from an exponential model under heteroscedasticity.
MP-MultDF is used to approximate the joint distribution for
MP testing methods.}
\label{fig:poa_het_lexpo2}
\end{figure}
\begin{figure}[h!]
\centering
\includegraphics[width=\textwidth]{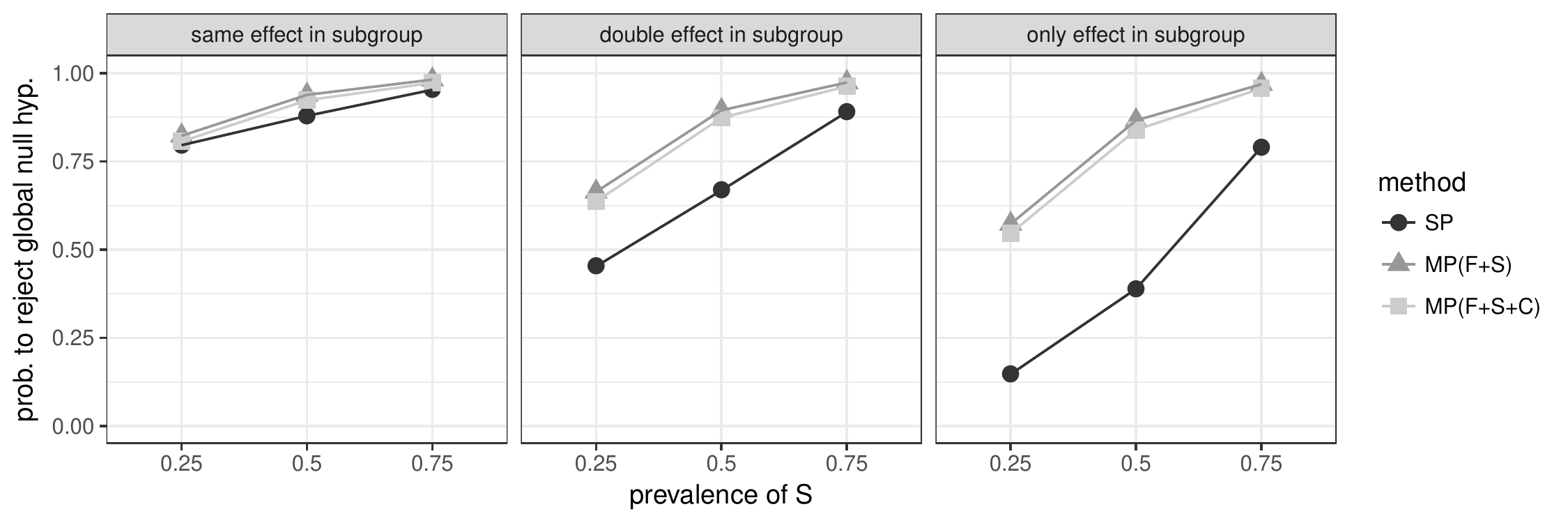}
\caption{Probability to reject the global null hypothesis for single population (SP) and
multi-population (MP) testing methods. Data
are generated from a logistic model under heteroscedasticity.
MP-MultDF is used to approximate the joint distribution for
MP testing methods.}
\label{fig:poa_het_log2}
\end{figure}
\begin{figure}[h!]
\centering
\includegraphics[width=\textwidth]{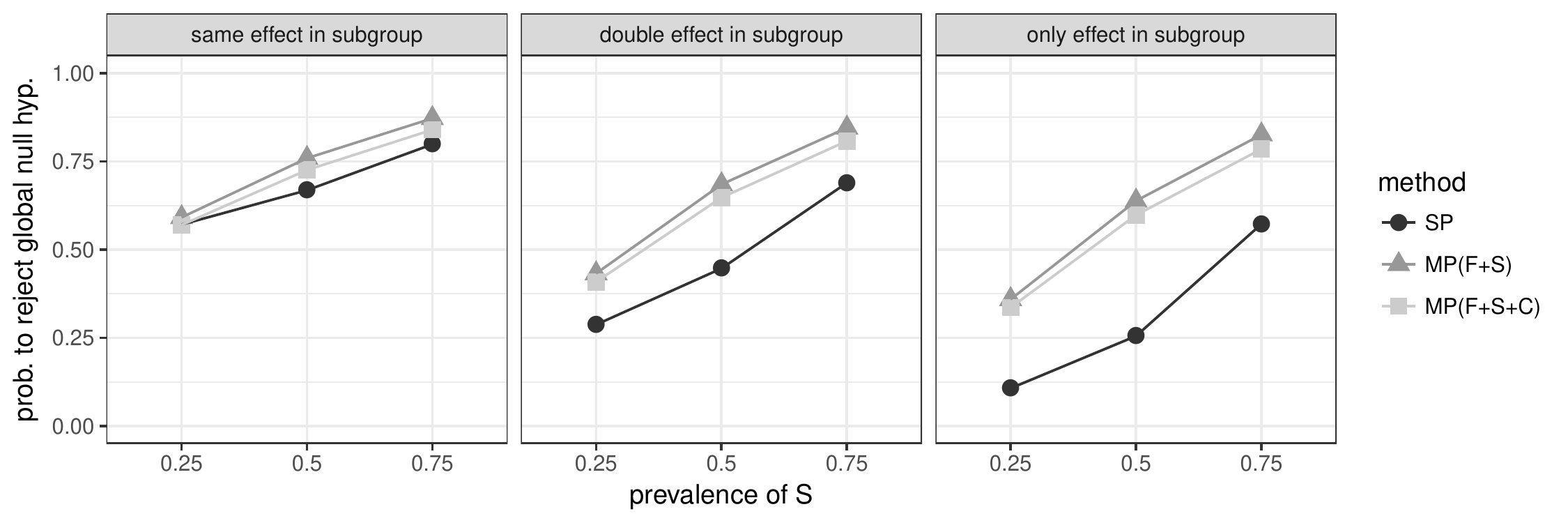}
\caption{Probability to reject the global null hypothesis for single population (SP) and
multi-population (MP) testing methods. Data
are generated from a quadratic model under heteroscedasticity.
MP-MultDF is used to approximate the joint distribution for
MP testing methods.}
\label{fig:poa_het_quad2}
\end{figure}

\clearpage

\end{document}